\documentclass[aps, prd, a4paper, floatfix,showpacs, preprintnumbers, superscriptaddress, nofootinbib, onecolumn]{revtex4}
\usepackage{amssymb,amsmath,amsfonts,amsbsy,graphicx,rotating,bbold}
\usepackage{color,microtype}
\usepackage{tabularx}
\usepackage{eucal}
\usepackage{epsfig}
\usepackage{float}

\usepackage{xcolor}
\usepackage{lipsum}
\usepackage{textcomp}
\usepackage{siunitx}
\usepackage{makecell}
\usepackage{lipsum}
\usepackage[T1]{fontenc}
\usepackage[utf8]{inputenc}
\usepackage[spanish,english]{babel}
\usepackage{listings}

\usepackage{hyperref}
\usepackage[labelsep=period]{caption}
\DeclareCaptionJustification{justified}{\leftskip=0pt \rightskip=0pt \parfillskip=0pt plus 1fil}
\hypersetup{
	colorlinks=true,
	linkcolor=blue,
	filecolor=red,   
	citecolor= red,
	urlcolor=magenta}
\makeatletter
\let\ams@underbrace=\underbrace
\def\underbrace#1_#2{``
	\setbox0=\hbox{$\displaystyle#1$}``
	\ams@underbrace{#1}_{\parbox[t]{\the\wd0}{#2}}``
}
\makeatother

\graphicspath{{Images/}}

\usepackage{subcaption}
\definecolor{vividviolet}{rgb}{0.62, 0.0, 1.0}
\definecolor{amaranth}{rgb}{0.9, 0.17, 0.31}
\definecolor{palatinateblue}{rgb}{0.15, 0.23, 0.89}
\definecolor{brightpink}{rgb}{1.0, 0.0, 0.5}
\definecolor{cornflowerblue}{rgb}{0.39, 0.58, 0.93}
\definecolor{deepcarminepink}{rgb}{0.94, 0.19, 0.22}
\definecolor{radicalred}{rgb}{1.0, 0.21, 0.37}
\def\beq{\begin{equation}}
\def\eeq{\end{equation}}
\usepackage[utf8]{inputenc}

\begin{document}

\title{Solving the H0 tension in $f(T)$ Gravity through Bayesian Machine Learning}

\author{Muhsin Aljaf}
\email{muhsinaljaf@oakland.edu}
\affiliation{Department of Physics, Oakland University, Rochester, Michigan, 48309, USA}
\affiliation{Department of Physics, College of Education, University of Garmian, Kurdistan Region-Iraq}

\author{Emilio Elizalde}
\email{elizalde@ice.csic.es }
\affiliation{
 Consejo Superior de Investigaciones Cient\'{\i}ficas, ICE/CSIC-IEEC,
Campus UAB, Carrer de Can Magrans s/n, 08193 Bellaterra (Barcelona) Spain }

\author{Martiros Khurshudyan}
\email{ khurshudyan@ice.csic.es, martiros.khurshudyan@csic.es }
\affiliation{
 Consejo Superior de Investigaciones Cient\'{\i}ficas, ICE/CSIC-IEEC,
Campus UAB, Carrer de Can Magrans s/n, 08193 Bellaterra (Barcelona) Spain }

\author{Kairat Myrzakulov}
\email{krmyrzakulov@gmail.com}
\affiliation{Eurasian National University, Nur-Sultan 010008, Kazakhstan}
\affiliation{Ratbay Myrzakulov Eurasian International Centre for Theoretical Physics, Nur-Sultan 010009, Kazakhstan}

\author{Aliya Zhadyranova}
\email{a.a.zhadyranova@gmail.com}
\affiliation{Eurasian National University, Nur-Sultan 010008, Kazakhstan}
\affiliation{Ratbay Myrzakulov Eurasian International Centre for Theoretical Physics, Nur-Sultan 010009, Kazakhstan}

\begin{abstract}
Bayesian Machine Learning~(BML) and strong lensing time delay~(SLTD) techniques are used in order to tackle the $H_{0}$ tension in $f(T)$ gravity. The power of  BML relies on employing a model-based generative process which already plays an important role in different domains of cosmology and astrophysics, being the present work a further proof of this. Three viable $f(T)$ models are considered: a power law, an exponential, and a squared exponential model. The learned constraints and respective results indicate that the exponential model, $f(T)=\alpha T_{0}\left(1-e^{-p T / T_{0}}\right)$, has the capability to solve the $H_{0}$  tension quite efficiently. The forecasting power and robustness of the method are shown by considering different redshift ranges and parameters for the lenses and sources involved. The lesson learned is that these values can strongly affect our understanding of the $H_{0}$ tension,  as it does happen in the case of the model considered. The resulting constraints of the learning method are eventually validated by using the observational Hubble data(OHD).

\end{abstract}

\maketitle

\section{Introduction}\label{intro}
The current standard model of cosmology, $\Lambda$CDM,  efficiently explains the evolution and content of the Universe by adding to its visible content two dark sectors: dark matter and dark energy. The first plays a crucial role in stabilizing galaxies and clusters, while the latter is necessary to describe the late-time acceleration of the Universe. However, even with these additions and remarkable efforts, the model still suffers from serious issues, such as the coincidence and the cosmological constant problems\cite{P1,P2}.

A rather new issue that reveals another trouble with the physical background of $\Lambda$CDM is the $H_{0}$ tension, to be discussed below. Essentially, there are two ways to tackle this new issue, widely discussed in the recent literature. One may just continue addressing the problems in the General Relativity~(GR) framework by adding new exotic forms of matter to the Universe's energy content or either build new gravitational theories beyond GR that could drive the accelerating expansion directly. In both approaches, the new models are bound to pass cosmological and astrophysical tests\cite{LCDM1, LCDM2, LCDM3, DE_start, DE_1, DE_2, DE_3, DE_4, DE_end, Sol1, Sol2} (See also references therein for related problems and models). 

In the context of the second approach, several modified theories have been proposed. For instance, the $f(R)$ theory is  the simplest extension of GR in which instead of Ricci scalar $R$ in the Einstein-Hilbert action one considers an arbitrary function $f(R)$\cite{F(R)1, F(R)2,F(R)3,F(R)4,F(R)5,F(R)6,F(R)7,F(R)8}. Another interesting modification is the $f(T)$ theory wherein the gravitational interaction is described by the Torsion $T$ instead of the curvature tensor. As a result, the Levi-Civita connection is replaced by the Weitzenböck connection in the underlying Riemann-Cartan spacetime. An important benefit of  $f(T)$ is that its field equations appear in the form of second-order differential equations, significantly reducing the mathematical difficulties of the models compared to $f(R)$ theories where the field equation leads to fourth-order differential equations. Moreover, the cosmological implications of $f(T)$ theories have already been manifested in several proposed models. These models are not only able to explain the current accelerated cosmic expansion but also provide alternatives to inflation. For all that, the $f(T)$ theory and its cosmological applications have attracted a lot of interest in the recent literature \cite{FT_3,AP1,AP2,AP3,AP4,AP5,AP6,AP7,AP8,cons1,cons2,cons3,cons4, f1,f2}.

We have already mentioned that it is essential to study and ensure that the crafted models pass cosmological and astrophysical tests, especially now that various observational missions are in operation, rendering lots of new data. Moreover, it is crucial to constrain the model parameters and learn proper consequences because the constraints on the background dynamics are essential for understanding the nature of (interacting) dark energy, structure formation, and future singularity problems. In this regard, developing and utilizing techniques that allow us to get reconstructions (including constraints) through a learning procedure in a model-independent way, e.g., directly from observational data, are of great importance. 

One of the popular and widely used examples of such techniques in cosmology is the Gaussian Processes\cite{GP1, GP2} which rely on a specific Machine Learning (ML) algorithm indicating how generally ML can be used in cosmology, astrophysics, or in any other field of science where data analysis is crucial. Generically, ML algorithms are data-hungry approaches requiring huge amounts of data to perform training and validation processes. They also carry some drawbacks that may cause catastrophic results in some cases\cite{GP3}. In other words, ML may become useless in specific situations, in particular, if the collected data have some inherent problem. Biasing is among the reasons that the ML approach may fail. Unfortunately, biasing is a substantial and sometimes unavoidable part of the data collecting process. It can originate from our particular understanding of reality or the model we use to represent reality. For instance, an intrinsic bias in flat $\Lambda$CDM has been pointed out by \cite{P3}. Generally, a bias can arise in many situations, including those associated with the reasoning under uncertainties actively studied in robotics, dynamical vehicles, and various autonomous system modeling. As a result, Over the years, researchers of computer science and other science fields have developed multiple methods to reduce bias and increase the robustness of ML algorithms. Among them, an interesting case for us, which will be applied in this paper, is Bayesian Machine Learning (BML). It uses model-based generative processes to improve the data problems, among others\footnote{We use BML as a tool to study cosmology and, for conciseness, have to omit various theoretical and technical details about it, including how it can be used in biased cases. We strongly suggest that readers interested in ML topics search on the web about recent developments and existing problems in this direction to gain more insight.}. A proper discussion about BML will appear below, in Sect.~\ref{method}, where it will be indicated how, in general, it can be used in cosmology.

The rest of this section is devoted to the formulation of the specific problem that motivates the present study. It is a relatively new one, known as the $H_{0}$ tension in the literature: a huge difference between the early-time measurements (e.g., Cosmic Microwave Background~(CMB) and Baryon acoustic oscillations~(BAO)) and late-time ones (e.g., Type Ia Supernovae (SNe Ia) and $\mathrm{H}(\mathrm{z}))$ of the value of the Hubble constant $H_{0}$. In particular, according to the Planck 2018 \cite{H01} results, in a flat $\Lambda$CDM model the value of the Hubble constant is $H_{0}=67.27 \pm0.60 \mathrm{~km} / \mathrm{s} / \mathrm{Mpc}$ at $1\sigma$ confidence level, while the SHOES Team estimated that $H_{0}=73.2\pm 1.3\mathrm{~km} / \mathrm{s} / \mathrm{Mpc}$ \cite{H02}, which exhibits a $4.14$$\sigma$ tension. 

This important tension motivated researchers to look for different solutions ranging from indications of new physics to possible hidden sources of systematic errors and biases in observational data \cite{H03, H04,H05,H06,H07,H08} (see references therein for other options to solve the $H_{0}$ tension). Indeed, to understand the source of such discrepancy that can challenge the $\Lambda$CDM model, other independent observational sources have been used to determine the value of $H_{0}$. For example, strong gravitational lensing systems are powerful and independent candidates for estimating the Hubble parameter and its current tension \cite{TD1}. The discovery of the first binary neutron star merging event, \textit{GW170817}, and the detection of an associated electromagnetic counterpart has made this possible, providing  an estimate for the Hubble constant of about $H_{0}=70_{-8}^{+12}\mathrm{~km} / \mathrm{s} / \mathrm{Mpc}$. Moreover, the analysis of six well-measured systems from the H0LiCOW lensing program \cite{HOLICOW} has provided abound on the Hubble constant of $73.3_{-1.8}^{+1.7}$ assuming a flat $\Lambda\mathrm{CDM}$ cosmology\cite{HOLICOW1}. Even though these constraints are weaker than those from  SNe Ia and CMB observations, it is expected to improve with the discovery of new merging events with an associated electromagnetic counterpart \cite{GW1, GW2, GW3, GW4, GW5, GW6}. Particularly with observations of the lensed systems from future surveys such as the Large Synoptic Survey Telescope (LSST) \cite{LSST1}, are expected to significantly improve the number of well-measured strongly lensed systems \cite{LSST2}. The increased number of observed lensed sources will also allow constraining non-standard cosmologies.

In strong gravitational lensing systems, the total time delay between two images (or two  gravitational-wave events), $i$ and $j$, is given by
\begin{equation}\label{SGLTD1}
	\begin{gathered}
		\Delta t_{i, j}=D_{\Delta \mathrm{t}}\left(1+z_{\mathrm{s}}\right) \Delta \phi_{i,j}, \\
		\Delta \phi_{i, j}=\left[\left(\boldsymbol{\theta}_{i}-\boldsymbol{\beta}\right)^{2} / 2-\psi\left(\boldsymbol{\theta}_{i}\right)-\left(\boldsymbol{\theta}_{j}-\boldsymbol{\beta}\right)^{2} / 2+\psi\left(\boldsymbol{\theta}_{j}\right)\right],
	\end{gathered}
\end{equation}
where $\Delta \phi_{i, j}$ is the difference between the Fermat potentials at different image angular positions $\boldsymbol{\theta}_{i}, \boldsymbol{\theta}_{j}$, and $\boldsymbol{\beta}$ denoting the source position, and $\psi$ being the lensing potential \cite{lens1,lens2,TD1}. 

On the other hand, the measured time delay between strongly lensed images $\Delta t_{i, j}$ combined with the redshifts of the lens $z_{\mathrm{s}}$ and the source $z_{\mathrm{s}}$, and the Fermat potential difference $\Delta \phi_{i, j}$ determined by lens mass distribution and image positions allow determining the time-delay distance $D_{\Delta t}$. This quantity, which is a combination of three angular diameter distances, reads
\begin{equation}\label{TB1}
	D_{\Delta t}=\left(1+z_{l}\right) \frac{D_{l} D_{s}}{D_{ls}},
\end{equation}
where $D_{l}, D_{s}$ and $D_{ls}$ stand for the angular diameter distances to the lens, the source, and between the lens and the source, respectively. In fact, Eq.(\ref{TB1}) is a very powerful relationship, combined with BML, which allows us to constrain cosmological models without relying on the physics or observations of the lensing model and Fermat potential. As a result, having only cosmological models and taking into account that in a flat Friedmann-Robertson-Walker~(FRW) Universe, the angular diameter distance reads 
\begin{equation}\label{TB2}
	D(z^{\prime})=\frac{1}{H_{0}(1+z)} \int_{0}^{z^{\prime}} \frac{d z^{\prime}}{E\left(z^{\prime},r\right)},
\end{equation}
where $E\left(z^{\prime},r\right)$ is defined as dimensionless Hubble parameter and it is  possible to learn the constraints on the model parameters embedded in it . Ideally, future observational data coming from lensed gravitational wave~(GW) signals together with their corresponding electromagnetic wave~(EM) will definitely provide us with some new and significant insights which may lead to an alleviation of the $H_{0}$ tension. Therefore, it is worth investigating their implications on the dark energy  and modified theories of gravity.
\begin{table}[h!]
	\centering
	\begin{tabular}{|c|c|c|c|c|c|}
		\hline
		$z$ & $H(z)$ & $\sigma_{H}$ & $z$ & $H(z)$ & $\sigma_{H}$ \\   
		$0.070$ & $69$ & $19.6$ & $0.4783$ & $80.9$ & $9$ \\         
		$0.090$ & $69$ & $12$ & $0.480$ & $97$ & $62$ \\
		$0.120$ & $68.6$ & $26.2$ &  $0.593$ & $104$ & $13$  \\
		$0.170$ & $83$ & $8$ & $0.680$ & $92$ & $8$  \\
		$0.179$ & $75$ & $4$ &  $0.781$ & $105$ & $12$ \\
		$0.199$ & $75$ & $5$ &  $0.875$ & $125$ & $17$ \\
		$0.200$ & $72.9$ & $29.6$ &  $0.880$ & $90$ & $40$ \\
		$0.270$ & $77$ & $14$ &  $0.900$ & $117$ & $23$ \\
		$0.280$ & $88.8$ & $36.6$ &  $1.037$ & $154$ & $20$ \\
		$0.352$ & $83$ & $14$ & $1.300$ & $168$ & $17$ \\
		$0.3802$ & $83$ & $13.5$ &  $1.363$ & $160$ & $33.6$ \\
		$0.400$ & $95$ & $17$ & $1.4307$ & $177$ & $18$ \\
		$0.4004$ & $77$ & $10.2$ & $1.530$ & $140$ & $14$ \\
		$0.4247$ & $87.1$ & $11.1$ & $1.750$ & $202$ & $40$ \\
		$0.44497$ & $92.8$ & $12.9$ & $1.965$ & $186.5$ & $50.4$ \\
		\hline  
		$0.24$ & $79.69$ & $2.65$ & $0.60$ & $87.9$ & $6.1$ \\
		$0.35$ & $84.4$ & $7$ &  $0.73$ & $97.3$ & $7.0$ \\
		$0.43$ & $86.45$ & $3.68$ &  $2.30$ & $224$ & $8$ \\
		$0.44$ & $82.6$ & $7.8$ &  $2.34$ & $222$ & $7$ \\
		$0.57$ & $92.4$ & $4.5$ &  $2.36$ & $226$ & $8$ \\ 
		\hline
	\end{tabular}
	\caption{Currently available observational Hubble data (OHD) used to validate the results of our analysis with BML. $H(z)$ and its uncertainty $\sigma_{H}$ are in units of $km/s/Mpc$. The upper panel corresponds to 30 samples deduced from the differential age method. The lower panel, to 10 samples obtained from the BAO method. See, for instance \cite{GP1} and references therein for more details about this data.}
	\label{tabledata}
\end{table}

In our study, we pursue two goals. First, we will use the advantages given by BML to constrain various $f(T)$ models using the physics of GW+EM systems. To our knowledge, this is the first time where BML and time-delay of GW+EM  systems have been both involved in studying $f(T)$ models. In this case, the generative process used in BML will be based on Eq.(\ref{TB1}) and Eq.(\ref{TB2}), thus establishing a direct link between cosmology and strong lensing time delay(SLTD). Our second goal will be to learn how the $H_{0}$ tension can be solved in $f(T)$ gravity. However, given specific aspects of the method, we have to limit ourselves to considering only three specific viable $f(T)$ models. We emphasize again, taking into account the specific aspects of our analysis, that the learned constraints will be validated using the observational Hubble data (OHD) obtained from cosmic chronometers and BAO data (see, for instance, \cite{GP1} and references therein). 

The validation of the BML results with OHD presented in our study demonstrates that it is reliable to learn possible biases between $H(z)$ and future SLTD data. Moreover, since BML uses a model-based generation process, we are here able, for forecasting purposes, to consider different situations to understand forthcoming data that could affect our understanding of the $H_{0}$ tension. In particular,  we should indicate that, by considering different redshift ranges and numbers for the lenses and sources, we have learned that future SLTD data may strongly affect our understanding of the $H_{0}$ tension. Hopefully, discussed predictions for the background dynamics can be validated by new missions and data in the near future, proving the forecasting and the robustness of the method used in our analysis.

This paper is organized as follows. In Sect.~(\ref{fTbd}), we provide a brief description of the cosmological dynamics in the frame of the $f(T)$ theory and introduce three specific $f(T)$ models to be constrained using BML. Sect.~(\ref{method}) provides the methodology of the BML approach used in our analysis. Finally, in Sect.~(\ref{results}),  our final results are presented. To finish, the conclusions that follow from the analysis are displayed in Sect.~(\ref{conclusions}).

\section{Theoretical framework and models}\label{fTbd}

In this section we briefly introduce the formalism of $f(T)$ gravity and its application in cosmology. Then we introduce three viable models that we will constrain in our study.

\subsection{$f(T)$ gravity }

 In $f(T)$ gravity the dynamical variables are the tetrad fields $e^A{ }_\mu$, where Greek indices correspond to the spacetime coordinates and Latin indices correspond to the tangent space coordinates. The tetrad fields $e^A{ }_\mu$ form an orthonormal basis in the tangent space at each point of the spacetime manifold. This implies that they satisfy the relation $g_{\mu \nu}=\eta_{A B} e^A{ }_\mu e^B{ }_\nu$, with $g_{\mu \nu}$ the spacetime metric and where $\eta_{A B}=(1,-1,-1,-1)$ is the tangent-space metric. 
 
The Weitzenböck connection in torsional gravity is defined as
\begin{equation}
  \hat{\Gamma}_{\mu \nu}^\lambda \equiv e_A^\lambda \partial_\nu e^A{ }_\mu=-e^A{ }_\mu \partial_\nu e_A^\lambda .  
\end{equation}

 This connection does not include the Riemann curvature but only a non-zero torsion, namely

\begin{equation}
T_{\mu \nu}^\lambda \equiv \hat{\Gamma}_{\nu \mu}^\lambda-\hat{\Gamma}_{\mu \nu}^\lambda=e_A^\lambda\left(\partial_\mu e_\nu^A-\partial_\nu e_\mu^A\right) .   
\end{equation}

Additionally, the torsion scalar is
\begin{equation}
T=S_\rho^{\mu \nu} T_{\mu \nu}^\rho,
\end{equation}

where
\begin{equation}
S_\rho^{\mu \nu} \equiv \frac{1}{2}\left(K_\rho^{\mu \nu}+\delta_\rho^\mu T_\alpha^{\alpha \nu}-\delta_\rho^\nu T_\alpha^{\alpha \mu}{ }_\alpha\right),
\end{equation}

with
\begin{equation}
  K_{\mu \nu}^\rho \equiv \frac{1}{2}\left(T_\mu^\rho \nu^\rho+T_{\nu^\rho}^\mu-T_{\mu \nu}^\rho\right).
\end{equation}

This theory is equivalent to general relativity at the level of equations of motion and and the generalized Lagrangian could be written as

\begin{equation}
S=\int d^4 x e \frac{M_P^2}{2}\left[T+f(T)+L_m\right]
\end{equation}
where $e=\operatorname{det}\left(e_\mu^A\right)=\sqrt{-g}, M_P$ is the Planck mass and $f(T)$ is the arbitrary function of torsion scalar $T$ (we use units where $c=1$ ). 

By varying the above action with respect to the tetrads, we obtain the field equations as
\begin{equation}
\begin{aligned}
&e^{-1} \partial_\nu\left(e e_A^\rho S_\rho^{\mu \nu}\right)\left[1+f_T\right]-e_A^\lambda T_{\nu \lambda}^\rho S_\rho^{\nu \mu}\left[1+f_T\right] \\
&+e_A^\rho S_\rho^{\mu \nu}\left(\partial_\nu T\right) f_{T T}+\frac{1}{4} e_A^\mu[T+f(T)] \\
&=4 \pi G e_A^\rho T(m)_\rho^\mu
\end{aligned}
\end{equation}
where $f_T \equiv \partial f(T) / \partial T, f_{T T} \equiv \partial^2 f(T) / \partial T^2$, and $T(m)_\rho^\mu$ is the matter energy-momentum tensor.

\subsection{Background Dynamics}

Concerning the background dynamics of the universe, we should study the cosmology of $f(T)$ gravity in the context of a homogeneous, isotropic, and spatially flat universe, characterized by $e_{\mu}^{A}=\operatorname{diag}(1, a, a, a)$, with the FLRW geometry described by 
\begin{equation}
d s^{2}=d t^{2}-a^{2}(t) \delta_{i j} d x^{i} d x^{j}.
\end{equation}
The Friedmann equations in this context become
\begin{equation}\label{fridmann1}
3 H^{2}=8 \pi G \rho_{m}-\frac{f}{2}+T f_{T},
\end{equation}
and
\begin{equation}\label{Hdot}
\dot{H}=-\frac{4 \pi G\left(\rho_{m}+P_{m}\right)}{1+f_{T}+2 T f_{T T}},
\end{equation}
with $H \equiv \frac{\dot{a}}{a}$ being the Hubble parameter, and $\rho_{dm}, P_{dm}$ being the energy density and pressure for cold dark matter, respectively. Accordingly,  we can define the energy density and pressure for dark energy as
\begin{equation}\label{rhoP}
\rho_{de} \equiv \frac{3}{8 \pi G}\left[-\frac{f}{6}+\frac{T f_{T}}{3}\right] \quad \text { and } \quad P_{de} \equiv \frac{1}{16 \pi G}\left[\frac{f-f_{T} T+2 T^{2} f_{T T}}{1+f_{T}+2 T f_{T T}}\right].
\end{equation}
Consequently, the dark energy equation of state can be written as
\begin{equation}
w_{de} \equiv \frac{P_{de}}{\rho_{de}}=\frac{f-f_{T} T+2 T^{2} f_{T T}}{\left(2 T f_{T}-f\right)\left(1+f_{T}+2 T f_{T T}\right)}.
\end{equation}
In the above discussed setup the cold dark matter will have its evolution dictated by the conservation of the energy-momentum tensor
\begin{equation}
\dot{\rho}_{dm}+3 H \rho_{dm}=0,
\end{equation}\label{cons1}
while the dark energy density will also follow the conservation equation
\begin{equation}\label{cons2}
\dot{\rho}_{de}+3H \rho_{de}\left(1+w_{de}\right)=0,
\end{equation}
with $\rho_{de}$ and $P_{de}$ defined by Eq.(\ref{rhoP}). Since $T=-6 H^{2}$, the normalized Hubble parameter $E(z)$ can be written as $E^{2}(z) \equiv \frac{H^{2}(z)}{H_{0}^{2}}=\frac{T(z)}{T_{0}}$, with $H_{0}$ being the present value of the Hubble parameter, and $T_{0}=-6 H_{0}^{2}$. It is worth to mention that, in our analysis, for convenience we re-write the Friedmann equation as 
\begin{equation}
E^{2}(z, r)=\Omega^{(0)}_{dm}(1+z)^{3}+\Omega^{(0)}_{de} y(z, r),
\end{equation}
with $y(z, r)$ being
\begin{equation}\label{distortion}
y(z, r) \equiv \frac{1}{6 H_{0}^{2} \Omega^{(0)}_{de}}\left[2 T f_{T}-f\right],
\end{equation}
and $\Omega^{(0)}_{de}$ being the dark energy density parameter today,
\begin{equation}
\Omega^{(0)}_{de} =1-\Omega^{(0)}_{dm},
\end{equation}
produced by the modifying $f(T)$ term. One can note the effect from the modified dynamics of teleparallel gravity is represented by the function $y(z, r)$ , in which $r$ corresponds to the free parameters of the specific model considered. The main characteristics of this function are that GR must be reproduced for some limit of parameter, while at the cosmological level, the concordance model $\Lambda$CDM can also be achieved ($y=1$).

\subsection{$f(T)$ Models}\label{model}
In this section we present three $f(T)$ models to be investigated in this work. The three selected functions have already been studied in the literature and are among the preferred ones by available data when compared to the $\Lambda$CDM model (see, for instance, \cite{cons1}). We will see how BML affects the predictions for each model while verifying the consistency with previous works. In what follows, we shall introduce the considered $f(T)$ forms and comment on their cosmological implications.

\begin{enumerate}
\item Power-law model

The first $f(T)$ model (hereafter $f_{1}$CDM) is the power-law model which reads as
\begin{equation}\label{eq:f1}
f_{1}= \alpha (-T)^{b},
\end{equation}
where $\alpha$ and $b$ are the two free parameters that can be related through
\begin{equation}\label{eq:f1alpha}
\alpha=\left(6 H_{0}^{2}\right)^{1-b}\frac{1-\Omega^{(0)}_{dm}}{2 b-1}.
\end{equation}
Taking $z=0, H(z=0)=H_{0}$ in Eq.(\ref{distortion}), the distortion factor becomes simply
\begin{equation}\label{eq:f1dist}
y(z, b)=E^{2 b}(z, b),
\end{equation}
and the Friedmann equation,
\begin{equation}\label{eq:f1Fried}
E^{2}(z, b)=\Omega_{m 0}(1+z)^{3}+\Omega^{(0)}_{de} E^{2 b}(z, b).
\end{equation}
We can easily see that $b=0$ reproduces the $\Lambda$CDM cosmology. This model gives a de-Sitter limit for $z=-1$, and deviations from the standard model are more evident for higher $|b|$. However, these deviations are generally small, as confirmed using numerical techniques.

\item Exponential model

The second $f(T)$ model (hereafter $f_{2}$CDM) is known as the exponential model, and reads 
\begin{equation}\label{eq:f2}
f_1(T)=\alpha T_0\left(1-e^{-p T / T_0}\right),
\end{equation}
where, again, $\alpha$ and $p$ are model parameters that can be related as
\begin{equation}\label{eq:f2alpha}
\alpha=\frac{\Omega^{(0)}_{de}}{1-(1+2 p) e^{-p}}.
\end{equation}
For this model the $\Lambda$CDM model is recovered when $p \rightarrow+\infty$ or equivalently $b \rightarrow 0^{+}$. Replacing $p=\frac{1}{b}$ in the above equation, the distortion becomes
\begin{equation}\label{eq:f2dist}
y(z, b)=\frac{1-\left(1+\frac{2 E^{2}}{b}\right) e^{-\frac{E^{2}}{b}}}{1-\left(1+\frac{2}{b}\right) e^{-\frac{1}{b}}}.
\end{equation}

and ,consequently, the Friedmann equation for this model becomes
\begin{equation}\label{eq:f2Fried}
E^{2}(z, b)=\Omega^{(0)}_{dm}(1+z)^{3}+\Omega^{(0)}_{de} \frac{1-\left(1+\frac{2 E^{2}}{b}\right) e^{-\frac{E^{2}}{b}}}{1-\left(1+\frac{2}{b}\right) e^{-\frac{1}{b}}}.
\end{equation}

\item The square-root exponential model

Finally, the third $f(T)$ model (hereafter $f_{3}$CDM) considered in this work is also of exponential form but has a different exponent, namely
\begin{equation}\label{eq:f3}
f_3(T)=\alpha T_0\left(1-e^{-p \sqrt{T / T_0}}\right), p=\frac{1}{b}
\end{equation}
where the $\alpha$ and $p$ parameters are related as
\begin{equation}\label{eq:f3alpha}
\alpha=\frac{\Omega^{(0)}_{de}}{1-(1+p) e^{-p}}.
\end{equation}

It is easy to see $\Lambda$CDM model is recovered when $p \rightarrow+\infty$ or equivalently $b\rightarrow 0^{+}$. Replacing $p=\frac{1}{b}$ to the above equation, the distortion factor becomes
\begin{equation}\label{eq:f3dist}
y(z, b)=\frac{1-\left(1+\frac{E}{b}\right) e^{-\frac{E}{b}}}{1-\left(1+\frac{1}{b}\right) e^{-\frac{1}{b}}},
\end{equation}
and Friedman equations becomes 
\begin{equation}\label{eq:f31}
E^{2}(z, b)=\Omega_{d m}^{(0)}(1+z)^{3}+\Omega_{d e}^{(0)} \frac{1-\left(1+\frac{E}{b}\right) e^{-\frac{E}{b}}}{1-\left(1+\frac{1}{b}\right) e^{-\frac{1}{b}}}.
\end{equation}

\end{enumerate}

\section{METHODOLOGY}\label{method}

In this section we review the building blocks of BML and discuss how it can be used to explore the parameter space of the background dynamics of the three $f(T)$ models introduced in the previous section.
\subsection{Bayesian Machine learning (BML)}
Constraining a cosmological scenario with observational data is the main tool in estimating whether the model is applicable or not. It also reduces the phenomenology step by step, revealing the acceptable scenario. Bayesian modeling and parameter inference with standard methods, Markov chain Monte Carlo (MCMC), play a central role in this chain. Here We will not discuss MCMC in great detail but provide a basic motivation to clarify why we need to look for alternative ways to constrain models. For this purpose, lets us  start with the Bayes theorem
\begin{equation}\label{eq:BT}
    P(\theta | \mathcal{D}) = \frac{P(\mathcal{D} |\theta) P(\theta)}{P(\mathcal{D})},
\end{equation}
where $P(\theta)$ is the prior belief on the parameter $\theta$ describing the model under consideration. $P(\mathcal{D}|\theta)$ is the likelihood that represents the probability of observing the data $\mathcal{D}$ given parameter $\theta$ or model (in our case would be the cosmological model). Finally, $P(\mathcal{D})$ is the marginal likelihood or model evidence. The Bayes theorem allows us to find the probability of a given model with $\theta$ parameters explaining given data $\mathcal{D}$. This probability is noted as $P(\theta|\mathcal{D})$ and actually a conditional probability. It should be mentioned that the marginal likelihood $P(\mathcal{D})$ is a useful quantity for model selection because it shows that the model will generate the data, irrespective of its parameter values. However the computation of the mentioned probabilities is impossible and this has led to the development and usage of alternative methods to overcome the intractable computational aspects. One of the alternative methods for performing Bayesian inference is the variational inference discussed below. It is considerably faster than MCMC techniques and does not suffer from convergence issues, making it very attractive for cosmological and astrophysical applications. 

Now, how can variational inference be helpful, and what is it? It is a helpful tool since it suggests solving an optimization problem by approximating the target probability density. The Kullback-Leibler (KL) divergence is used as a measure of such proximity \cite{K.L.}. In our case, the target probability density would be the Bayesian posterior which allows us to constrain the model parameters. For this purpose, the first step is finding or proposing a family of densities $\mathcal{Q}$ and then finding the member of that family $q(\theta) \in \mathcal{Q}$ which is the closest one to the target probability density. This member is   known as the variational posterior that minimizes the KL  divergence to the exact posterior, that is  
\begin{equation}
q^{*}(\theta)=\underset{q(\theta) \in \mathcal{Q}}{\arg \min } \mathrm{KL}(q(\theta)|p(\theta|\mathcal{D})),
\end{equation}
 where $\theta$ is the latent variable to measure of such proximity, and the KL divergence is defined as
\begin{equation}\label{eq:KL1}
\mathrm{KL}(q(\theta) \| p(\theta \mid  \mathcal{D}))=\mathbb{E}_{q(\theta)}[\log q(\theta)]-\mathbb{E}_{q(\theta)}[\log p(\theta \mid  \mathcal{D})].
\end{equation}
Using Bayes theorem, we can rewrite the above KL divergence as   
\begin{equation}\label{eq:KL2}
\mathrm{KL}(q(\theta) \| p(\theta \mid  \mathcal{D}))=\log p( \mathcal{D})+\mathbb{E}_{q(\theta)}[\log q(\theta)]-\mathbb{E}_{q(\theta)}[\log p( \mathcal{D}, \theta)].
\end{equation}

It can be noticed from the above equation that in order to  minimize the above KL divergence term, one needs to minimize the second and third terms in Eq.(\ref{eq:KL2}). Now, expanding the joint likelihood $p(\mathcal{D}, \theta)$ in Eq.(\ref{eq:KL2}) the variational  lower bound can be rewritten as \cite{ELBO} 
\begin{equation}\label{eq:ELBO}
\operatorname{ELBO}(q(\theta))=\mathbb{E}_{q(\theta)}[\log p( \mathcal{D} \mid \theta)]-\mathrm{KL}(q(\theta) \| p(\theta)).
\end{equation}
 The first term in the above equation is a sort of data fit term maximizing the likelihood of the observational data. In contrast, the second term is KL-divergence between the variational distribution and the prior. It can be interpreted as a regularization term that ensures that the variational distribution does not become too complex, potentially leading to over-fitting. That being said, we have a fitting tool to control the process and efficiently avoid computation problems by increasing or decreasing the contribution of the first two terms in Eq.(\ref{eq:KL1}). There is interesting and valuable information on this topic which can be found in \cite{51,52,53,54}, to mention a few references. 

However, the above discussion does not answer how we can find the approximation for the target probability density. An easy option could be to guess it for a simple model where the Bayes inference is also tractable. In practice, this is not an option. The other option is to learn it from the model directly (in our case, directly from the crafted cosmological model), using Neural Networks (NN). In this case, both terms in Eq.(\ref{eq:ELBO}) can be interpreted from a new perspective, reducing to the initial personal belief and the generated model belief, respectively. It is a convenient approach that allows overcoming various data-related issues because, in this case, even low-quality data can be used at the end to validate the learned results. This idea and its success represent years of development, allowing to transfer the whole subject to another level. Eventually, we should mention that as a learning method, we use deep probabilistic learning, a type of deep learning accounting for uncertainties in the model, initial belief, belief update and deep neural networks. This approach provides the adequate groundwork to output reliable estimations for many ML tasks. 

\subsection{Implementation of BML}

We will make use of the probabilistic programming package \textit{PyMC3} \cite{PyMC3}, which uses the deep learning library Theano, that is, a deep learning python-based library, providing cutting edge inference algorithms to define the physical model, to perform variational inference and to build the posterior distribution. We found that the \textit{PyMC3} public library is enriched with some excellent examples demonstrating how the learning process can be established and the probability distributions can be learned; therefore, we excluded any specific discussion on the mathematical framework behind ML algorithms and BML. We strongly suggest that readers interested in exploring BML and variational inference follow the examples and discussions provided in the \textit{PyMC3} manual.

Now, let us discuss how we should understand the above discussion allowing us to integrate BML and variational inference to learn the constraints on the crafted cosmological model. To be short, in our analysis using PyMC3, we have followed the steps:
\begin{enumerate}
\item We define our cosmological model and the observable that would be generated, and hence we establish the elements of the so-called generative process. In this paper, this process is based on Eqs.(\ref{TB1}) and (\ref{TB2}).

\item We treat the data obtained from the generative process as data in the following sense. It is very important to understand the meaning of this step. In particular, having generated the so-called data, we now generate probability distributions showing how a given cosmological model can explain the data. In this way, we significantly reduce the complexity of the problem because the family of probabilities approximating the final posterior will directly depend only on priors. To notice dependency, we need to consider the meaning of the right-hand side of the Bayes theorem, Eq.(\ref{eq:BT}). This is a good starting point as we will see below.

\item Finally, we run the learning algorithm to get a new distribution over the model parameters and update our prior beliefs imposed on the cosmological model parameters.
\end{enumerate}

Generating the learning process and the direction of the learning is always controlled by KL divergence. Since we have involved probabilistic programming, after enough generated probabilistic distributions, we expect to learn the asymptotically correct form for the posterior distribution allowing us to infer the constraints.  

In our analysis we considered several different scenarios where data have been generated, which should be understood in the context of the above discussion, to cover different redshift ranges for both lens and sources and different numbers of the lenses and sources. The distribution for the lens $z_{l}$ and source $z_{s}$ considered in this paper can be found in Table \ref{ranges}.
\begin{table}[h!]
\centering
\begin{tabular}{|c|c|}
\hline $\text{Lens distribution} (z_{l})$ & $\text{Source distribution}(z_{s})$ \\
\hline$(0.1,1.2)$ & {$(0.3,1.7)$} \\
\hline$(0.1,1.5)$ & {$[0.3,1.7)$} \\
\hline$(0.1,2.0)$ & {$(0.3,2.5)$} \\
\hline$(0.1,2.4)$ & {$[0.3,2.5)$} \\
\hline
\end{tabular}
\caption{The redshift distribution prior on the lens and source of GW+EM lensed systems considered in this paper. The initial belief used as an input to start the learning is the $\Lambda$CDM model.}
\label{ranges}
\end{table}
To end this section, we would like to mention that we cover lenses and sources distributed over both low and high redshifts. This is because the observational data for the cosmic history of the Universe are available at low redshift ranges and can be used to validate the learned results obtained from BML. On the other hand, we consider high redshift ranges for forecasting reasons; however, the complete validation of the results will have to wait for the near future, when observations of higher redshift data actually become available. Indeed the validation of the learned results is based on the expansion rate data presented in Table \ref{tabledata}. Moreover, we need to stress that our initial belief used as an input is the $\Lambda$CDM model. Even if we start from different initial beliefs, our learning procedure asymptotically converges to the results discussed in this paper.

\section{Learned constraints on the model parameters}\label{results}

In this section, we present the learned constraints on the $f(T)$ cosmological parameters obtained following the procedure described in Sect.~\ref{method}. For the sake of convenience, we provide our results in three subsections.

\subsection{$f_{1}$CDM}

The first case corresponds to the $f_{1}$CDM model given by Eqs.~(\ref{eq:f1}) and (\ref{eq:f1Fried}) when the generative process for BML has been organized using Eqs.~(\ref{TB1}) and (\ref{TB2}). Moreover, for all cases discussed below, flat priors as $H_{0} \in [64,78]$, $\Omega^{(0)}_{dm} \in [0.2,0.4]$ and $b \in [-0.1,0.1]$ have been imposed, respectively. Also we need to mention that $N$ lenses (and respective sources) for a given redshift range are distributed uniformly according to the intervals given in Table \ref{ranges}. From our analysis, we have learned that:
\begin{itemize}
    \item The best fit values and $1\sigma$ errors of the model parameters when $z_{l}\in[0.1,1.2]$ and $z_{s}\in[0.3,1.7]$ for $N _{lens}= 50$ are $\Omega^{(0)}_{dm} = 0.28 \pm 0.012$, $H_{0} = 69.61 \pm 0.141$  km/s/Mpc and $b = 0.00197 \pm 0.005$.
    
    \item On the other hand, when $z_{l}\in[0.1,1.2]$ and $z_{s}\in[0.3,1.7]$ with $N _{lens}= 100$ the best fit values of the model parameters are found to be $\Omega^{(0)}_{dm} = 0.29 \pm 0.012$, $H_{0} = 68.81 \pm 0.145$ km/s/Mpc and $b = 0.003 \pm 0.005$. 
    
    \item Moreover, for the other two cases when $z_{l}\in[0.1,1.5]$ and $z_{s}\in[0.3,1.7]$, and $z_{l}\in[0.1,2.0]$ and $z_{s}\in[0.3,2.5]$ (in both cases $N _{lens}= 100$), we have found that $\Omega^{(0)}_{dm} = 0.297 \pm 0.011$, $H_{0} = 68.84 \pm 0.146$  km/s/Mpc and $b = 0.0043 \pm 0.0051$, and $\Omega^{(0)}_{dm} = 0.328 \pm 0.012$, $H_{0} = 69.06 \pm 0.175$  km/s/Mpc and $b = 0.0071 \pm 0.0051$, respectively. 
    
    \item Finally, we found $\Omega^{(0)}_{dm} = 0.302 \pm 0.0105$, $H_{0} = 65.56 \pm 0.134$ km/s/Mpc and $b = 0.00013 \pm 0.0005$, when $z_{l}\in[0.1,2.4]$ and $z_{s}\in[0.3,2.5]$ for $N _{lens}= 100$.
    
\end{itemize}

\begin{table}[h!]
  \centering
    \begin{tabular}{ | c | c | c | c |  c |  c |  p{2cm} |}
    \hline
    
  $f_{1}$CDM: $f(T) = \alpha (-T)^{b}$ & $N_{lens}$ & $\Omega^{(0)}_{dm}$ & $H_{0}$ & $b$  \\
      \hline
      
 when  $z_{l}\in[0.1,1.2]$ and $z_{s}\in[0.3,1.7]$ & $50$ & $0.28 \pm 0.012$ & $69.61 \pm 0.141$ & $0.00197 \pm 0.005$\\
          \hline
          
when  $z_{l}\in[0.1,1.2]$ and $z_{s}\in[0.3,1.7]$  & $100$ & $0.29 \pm 0.012$ & $68.81 \pm 0.145$ & $0.003 \pm 0.005$ \\
         \hline

when $z_{l}\in[0.1,1.5]$ and $z_{s}\in[0.3,1.7]$ & $100$ & $0.297 \pm 0.011$  & $68.84 \pm 0.146$ & $0.0043 \pm 0.0051$ \\
         \hline
         
 when $z_{l}\in[0.1,2.0]$ and $z_{s}\in[0.3,2.5]$ &  $100$  & $0.328 \pm 0.012$ & $69.06 \pm 0.175$ & $0.0071 \pm 0.0051$ \\  
 
     \hline 
     
 when $z_{l}\in[0.1,2.4]$ and $z_{s}\in[0.3,2.5]$ &  $100$  & $0.302 \pm 0.0105$ & $65.56 \pm 0.134$ & $0.00013 \pm 0.0005$ \\  
 
     \hline      
    
    \end{tabular}
\caption{The best fit values and $1\sigma$ errors estimated for $f_{1}$CDM model given by Eqs.~(\ref{eq:f1}) and (\ref{eq:f1Fried}) when the model based generation process is carried out on Eqs.~(\ref{TB1}) and (\ref{TB2}). The flat priors as $H_{0} \in [64,78]$, $\Omega^{(0)}_{dm} \in [0.2,0.4]$ and $b \in [-0.1,0.1]$ have been imposed and used in the generative process, respectively. Note that $H_{0}$ is measured in units of $ \mathrm{km} / \mathrm{s} / \mathrm{Mpc}$.}
  \label{tab:Table1}
\end{table} 

\begin{figure}[h!]
\centering
 \includegraphics[width=95 mm]{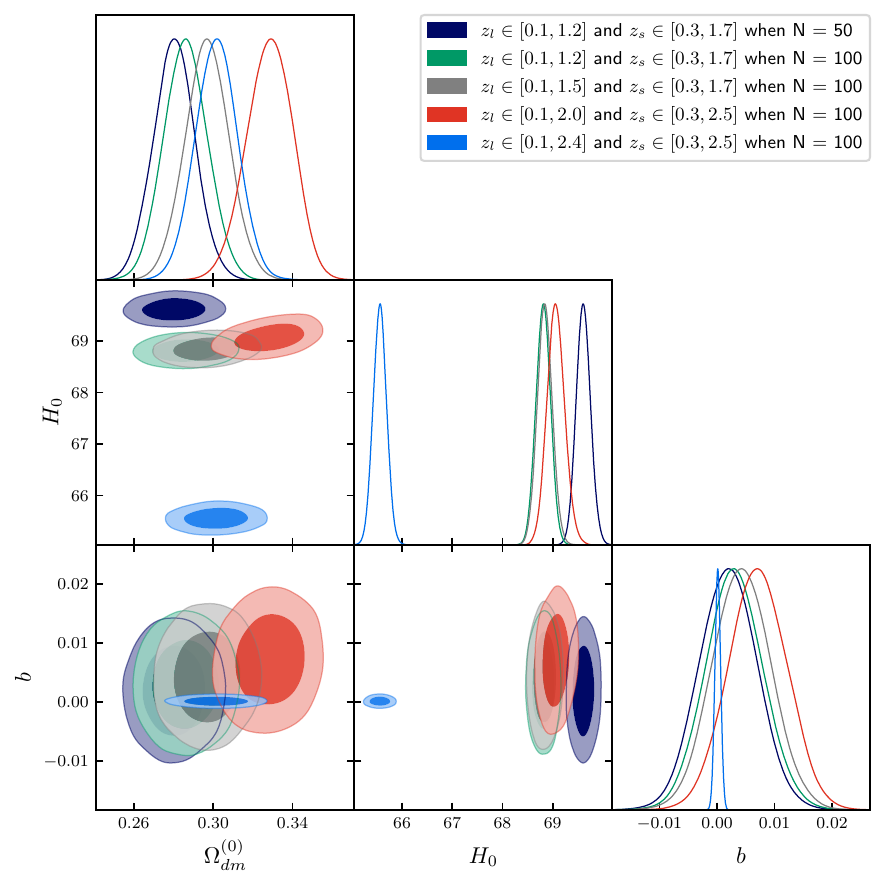}
\caption{$1\sigma$ and $2\sigma$ confidence-level contour plots for the cosmological parameters and the  parameter $b$ for $f_{1}$CDM model using the SLTD simulated data obtained from the generative process based on Eqs. (\ref{TB1}) and (\ref{TB2}).  Each contour color stands for lenses and sources distributed over a specific redshift range with lens number $N_{\text{lens}}=50$ (navy contour) and $N_{\text{lens}}=100$ (green, gray, red, and blue contours), respectively. The flat priors as $H_{0} \in [64,78]$, $\Omega^{(0)}_{dm} \in [0.2,0.4]$ and $b \in [-0.1,0.1]$ have been imposed  and used in the generative process. Our initial belief used as an input is the $\Lambda$CDM model.}
 \label{fig:Fig1_0}
\end{figure}

\begin{figure}[h!]
 \centering
     \begin{subfigure}[]{0.44\textwidth}
         \centering
         \includegraphics[width=\textwidth]{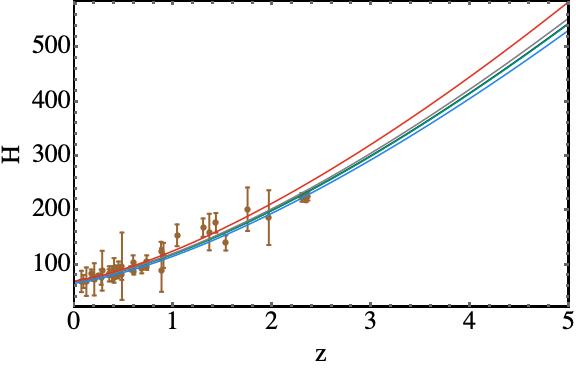}
         \caption{}
         \label{}
     \end{subfigure}
       \hskip -1ex
     \begin{subfigure}[]{0.44\textwidth}
         \centering
         \includegraphics[width=\textwidth]{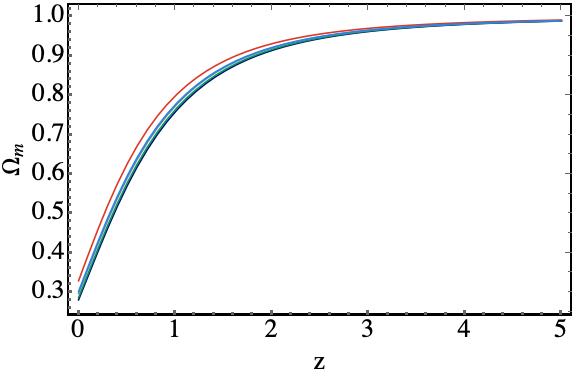}
         \caption{}
         \label{}
     \end{subfigure}
       \hskip -1ex
     \begin{subfigure}[]{0.44\textwidth}
         \centering
         \includegraphics[width=\textwidth]{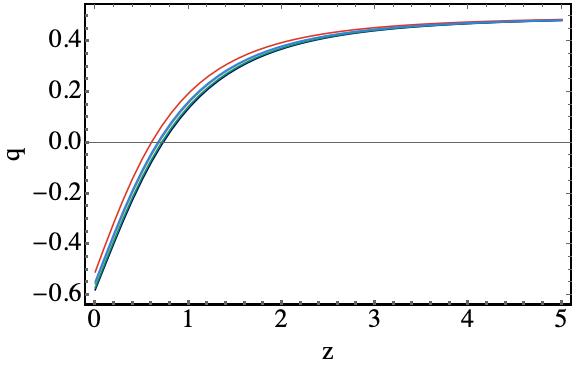}
         \caption{}
         \label{}
     \end{subfigure}
           \hskip -1ex
     \begin{subfigure}[]{0.44\textwidth}
         \centering
         \includegraphics[width=\textwidth]{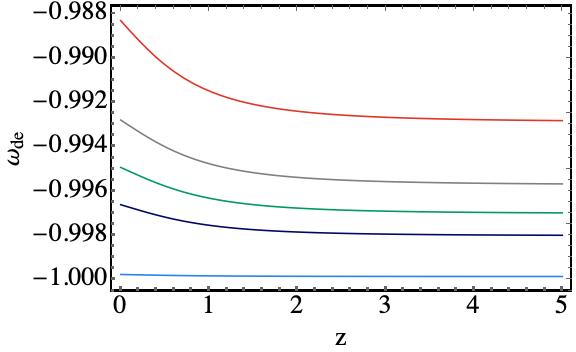}
         \caption{}
         \label{}
     \end{subfigure}
\caption{BML predictions for the redshift evolution of the Hubble parameter $H$, dark matter abundance $\Omega_{m}$, deceleration parameter $q$,  and equation of state parameter $\omega_{de} = \frac{p_{de}}{\rho_{de}}$ for the best fit values of the model parameters of $f_{1}$CDM model presented in Table \ref{tab:Table1}. Each color line stands for lenses and sources distributed over a specific redshift range with lens number $N_{\text{lens}}=50$ (navy curve) and $N_{\text{lens}}=100$, green, gray, red, and blue curves, respectively.}
 \label{fig:Fig1_1}
\end{figure}

A compact summary of the learned results can be found in Table \ref{tab:Table1}, while Fig. \ref{fig:Fig1_0} represents the $1\sigma$ and $2\sigma$ contour map of $f_{1}$CDM. It is easy to see that BML imposed very tight constraints on the parameters. We note that the parameter $b$, which determines the deviation from the $\Lambda$CDM model, is close to zero in all cases, indicating that according to SLTD measurements, the $f_{1}$CDM model most likely does not deviate from the $\Lambda$CDM model. This is not surprising because similar conclusions have already been achieved in the literature. However, we need to stress that this is the first indication that the developed pipeline is robust and allows us to learn previously known results from a completely different setup, which is impossible to reproduce with classical methods used in cosmology. 
As in any other ML algorithm, we also need to validate our BML learned results, and for this purpose, we use available OHD as discussed already. In our opinion, it is reasonable to follow this particular way of validating the learned results because we aimed to learn how to solve the $H_{0}$ tension in this particular model.

The graphical results of the validation process can be found in panel (a) of Fig. \ref{fig:Fig1_1} where we compare the redshift evolution of the Hubble parameter predicted by BML with OHD from cosmic chronometers and BAO. We also study the dark matter abundance $\Omega^{(0)}_{dm}$, deceleration parameter $q$, and equation of state parameter $\omega_{de}$ for a better understanding of the background dynamics which can be found in (b), (c), and (d) panels, respectively. We keep the same convention for the navy, green, grey, red, and blue curves as the legends in Fig. \ref{fig:Fig1_0}, and the dots correspond to the $40$ data points representing available OHD to be found in Table \ref{tabledata}. 

We notice that, according to the mean of the learned results, the redshift evolution of the Hubble function matches the OHD at low redshifts perfectly, but some tension arises at high redshifts. This is an interesting warning of the learning method, which should be kept under control in the future analysis of this model with strong lensing time delay data. Additionally, from panel (c) of Fig. \ref{fig:Fig1_1} we observe a good phase transition between a decelerating and an accelerating phase in all studied cases. On the other hand, the panel (d) of Fig. \ref{fig:Fig1_1} shows that the model behaves like the cosmological constant as $\omega_{de} \approx-1$. However, a deviation from the cosmological constant is also expected to observe. Interestingly, this result, in its turn, clearly indicates support for dynamical dark energy models.

In conclusion,  we see that the $f_{1}$CDM model is not able to solve the $H_{0}$ tension, and new observational data with significantly increased lens-source numbers will eventually disfavor the model. The same claim, with high fidelity according to learned results, can be said even when the systems are observed beyond currently available redshift ranges. Eventually, another significant result we learned, which should be tackled in the future properly, is that the tension between OHD and SLTD data at high redshifts should be considered seriously.  

\subsection{$f_{2}$CDM}

The second model to be considered is  $f_{2}$CDM given by Eqs.~(\ref{eq:f2}) and (\ref{eq:f2Fried}). In this case, the generative process for BML has been also organized following Eqs.~(\ref{TB1}) and (\ref{TB2}). During the study of this model we have learned  the best fit values of the model parameters with their $1\sigma$ errors. In particular, we found:
\begin{itemize}
    \item When $z_{l}\in[0.1,1.2]$ and $z_{s}\in[0.3,1.7]$ for $N _{lens}= 50$ the best fit values and $1\sigma$ errors to be $\Omega^{(0)}_{dm} = 0.277 \pm 0.01$, $H_{0} = 73.59^{+0.174}_{-0.165}$  km/s/Mpc and $p = 5.88^{+0.29}_{-0.34}$, respectively.
    
    \item On the other hand, when $z_{l}\in[0.1,1.2]$ and $z_{s}\in[0.3,1.7]$ with $N _{lens}= 100$ the best fit values of the model parameters are found to be $\Omega^{(0)}_{dm} = 0.28 \pm 0.01$, $H_{0} = 74.68 \pm 0.148$  km/s/Mpc and $p = 6.01 \pm 0.256$.
    
    \item Moreover, for the other two cases when $z_{l}\in[0.1,1.5]$ and $z_{s}\in[0.3,1.7]$, and $z_{l}\in[0.1,2.0]$ and $z_{s}\in[0.3,2.5]$ (in both cases $N _{lens}= 100$), we have found that $\Omega^{(0)}_{dm} = 0.279 \pm 0.0097$, $H_{0} = 74.90 \pm 0.135$  km/s/Mpc and $p = 5.26 \pm 0.179$ with  $\Omega^{(0)}_{dm} = 0.276 \pm 0.01$, $H_{0} = 74.64 \pm 0.145$  km/s/Mpc and $p = 5.4 \pm 0.135$, respectively.
    
    \item Finally, we find that $\Omega^{(0)}_{dm} = 0.275 \pm 0.01$, $H_{0} = 74.41^{+0.183}_{-0.178}$  km/s/Mpc and $p = 5.67^{+0.175}_{-0.17}$. This is obtained for the case when $z_{l}\in[0.1,2.4]$ and $z_{s}\in[0.3,2.5]$ for $N _{lens}= 100$.
\end{itemize}

For all cases discussed above, flat priors as $H_{0} \in [64,78]$, $\Omega^{(0)}_{dm} \in [0.2,0.4]$ and $p \in [-10,10]$ have been imposed. Moreover, $N$ lenses (and sources respectively) for all the considered cases are distributed uniformly as in the case of $f_{1}$CDM model discussed in the previous section. 
\begin{table}[h!]
  \centering
\begin{tabular}{ | c | c | c | c |  c |  c |  p{2cm} |}
    \hline
    
  $f_{2}$CDM: $f(T)=\alpha T_{0}\left(1-e^{-p T / T_{0}}\right)$ & $N_{lens}$ & $\Omega^{(0)}_{dm}$ & $H_{0}$ & $p$  \\
      \hline
      
 when  $z_{l}\in[0.1,1.2]$ and $z_{s}\in[0.3,1.7]$ & $50$ & $0.277 \pm 0.01$ & $73.59^{+0.174}_{-0.165}$ & $5.88^{+0.29}_{-0.34}$\\
          \hline
          
when  $z_{l}\in[0.1,1.2]$ and $z_{s}\in[0.3,1.7]$  & $100$ & $0.28 \pm 0.01$ & $74.68 \pm 0.148$ & $6.01 \pm 0.256$ \\
         \hline

when $z_{l}\in[0.1,1.5]$ and $z_{s}\in[0.3,1.7]$ & $100$ & $0.279 \pm 0.0097$  & $74.90 \pm 0.135$  & $5.26 \pm 0.179$ \\
         \hline
         
 when $z_{l}\in[0.1,2.0]$ and $z_{s}\in[0.3,2.5]$ &  $100$  & $0.276 \pm 0.01$ & $74.64 \pm 0.145$ & $5.4 \pm 0.135$ \\  
 
     \hline 
     
 when $z_{l}\in[0.1,2.4]$ and $z_{s}\in[0.3,2.5]$ &  $100$  & $0.275 \pm 0.01$ & $74.41^{+0.183}_{-0.178}$  & $5.67^{+0.175}_{-0.17}$ \\  
 
     \hline      
    
    \end{tabular}
\caption{Best fit values and $1\sigma$ errors estimated for $f_{2}$CDM given by Eqs.~(\ref{eq:f2}) and (\ref{eq:f2Fried}), when the model generation process is based on Eqs.~(\ref{TB1}) and (\ref{TB2}). The flat priors as $H_{0} \in [64,78]$, $\Omega^{(0)}_{dm} \in [0.2,0.4]$ and $p \in [-10,10]$ have been imposed and used in the generative process, respectively. $H_{0}$ is measured in  units of $ \mathrm{km} / \mathrm{s} / \mathrm{Mpc}$.}
  \label{tab:Table2}
\end{table} 

\begin{figure}[h!]
\centering
\includegraphics[width=95 mm]{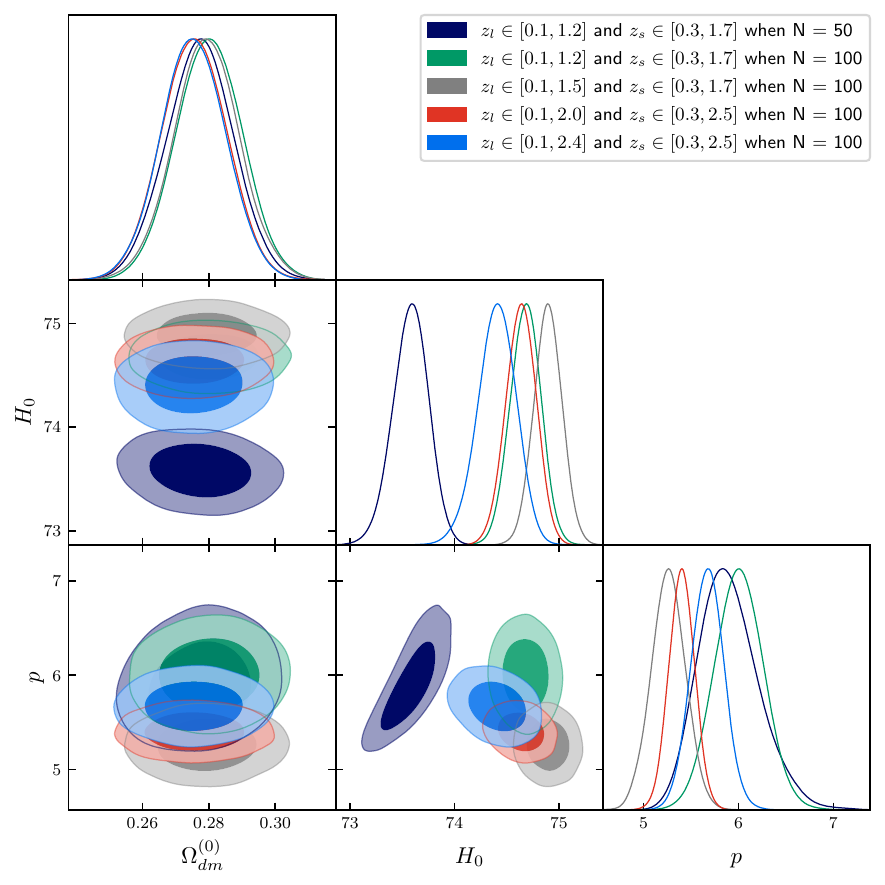}
\caption{The $1\sigma$ and $2\sigma$ confidence-level contour plots for the $f_{2}$CDM model, using the of SLTD  simulated datasets obtained from the generative process based on Eqs.~(\ref{TB1}) and (\ref{TB2}).  Each contour color stands for lenses and sources distributed over a specific redshift range, with lens number $N_{\text{lens}}=50$ (navy contour) and $N_{\text{lens}}=100$, green, gray, red, and blue contours, respectively. The flat priors as $H_{0} \in [64,78]$, $\Omega^{(0)}_{dm} \in [0.2,0.4]$ and $p \in [-10,10]$ have been imposed and used in the generative process. Our initial belief, used as an input, is the $\Lambda$CDM model.}
\label{fig:Fig2_0}
\end{figure}

\begin{figure}[h!]
\centering
     \begin{subfigure}[b]{0.44\textwidth}
         \centering
         \includegraphics[width=\textwidth]{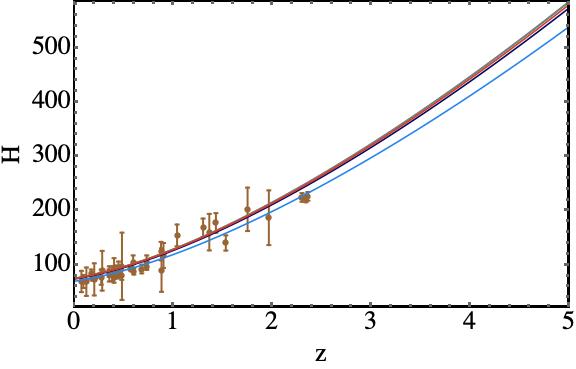}
         \caption{}
         \label{}
     \end{subfigure}
     \hskip -1ex
     \begin{subfigure}[b]{0.44\textwidth}
         \centering
         \includegraphics[width=\textwidth]{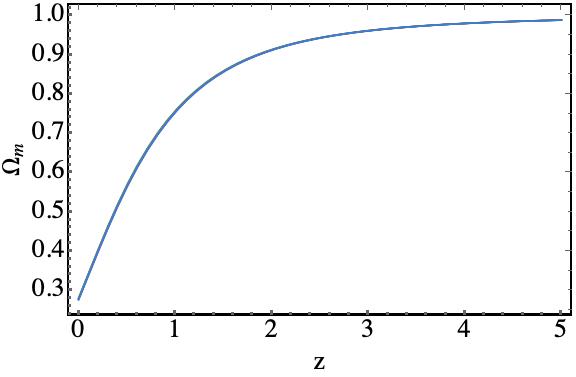}
         \caption{}
         \label{}
     \end{subfigure}
     \hskip -1ex
     \begin{subfigure}[b]{0.44\textwidth}
         \centering
         \includegraphics[width=\textwidth]{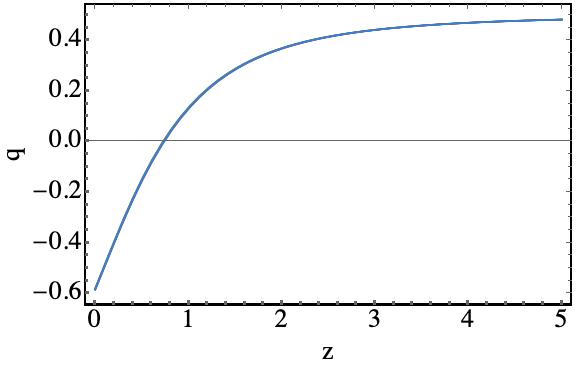}
         \caption{}
         \label{}
     \end{subfigure}
         \hskip -1ex
     \begin{subfigure}[b]{0.44\textwidth}
         \centering
         \includegraphics[width=\textwidth]{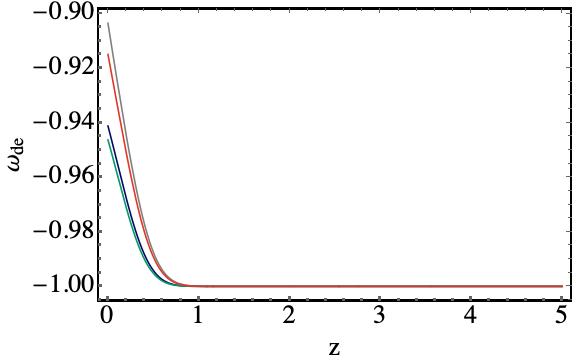}
         \caption{}
         \label{}
     \end{subfigure}
\caption{The BML predictions for the redshift evolution of the Hubble parameter $H$, dark matter abundance $\Omega_{m}$, deceleration parameter $q$,  and equation of state parameter $\omega_{de} = \frac{p_{de}}{\rho_{de}}$ for the best fit values of the model parameters of $f_{2}$CDM presented in Table \ref{tab:Table2}. Each  color line stands for lenses and sources distributed over a specific redshift range with lens number $N_{\text{lens}}=50$ (navy curve) and $N_{\text{lens}}=100$, green, gray, red, and blue curves, respectively.}
 \label{fig:Fig2_1}
\end{figure}

The learned results are summarized in Table \ref{tab:Table2} and Fig. \ref{fig:Fig2_0} shows the learned contour plots of the model parameters. The BML again imposes tight constraints on the cosmological parameters for all the considered cases. The learned results indicate that the model at hand deviates more from the standard cosmological model| due to the relatively significant value of the parameter $b=1/p$ compared to the previous model. Moreover, we provide validation for our BML results in the panel (a) of Fig. \ref{fig:Fig2_1} where we compare the learned redshift evolution of the Hubble parameter with available OHD. The same Fig. \ref{fig:Fig2_1}, but for the panels (b), (c) and (d), provides the graphical behavior of dark matter abundance $\Omega_{m}$, the deceleration parameter $q$,and equation of state parameter $\omega_{de}$, respectively, taking into account learned mean values of the model parameters. The same convention for the navy, green, grey, red, and blue curves as the legends in Fig. \ref{fig:Fig2_0} is used. 

Apparently, the dots  in panel (a) of Fig. \ref{fig:Fig2_1} represent again the $40$ data points from Table \ref{tabledata}. We note that the BML prediction for the redshifts evolution of the Hubble function matches the observational data at low redshifts, but we observe some tension at high redshifts. On the other hand, in the panel (c) of Fig. \ref{fig:Fig2_1}, we see a phase transition between a decelerating and an accelerating Universe phase in all considered cases. Moreover, panel (d)  shows that the model behaves like the cosmological constant, $\omega_{de}=-1$, at $z\geq 0.75$ and the quintessence dark energy, $\omega_{de}>-1$, at $z<0.75$. The motioned behaviour holds for all cases considered, indicating an almost linearly increasing functional form for $\omega_{d}$ for $z \in [0,0.75]$. 

Interestingly, we can say that the model can alleviate the $H_{0}$ tension because we have a significant deviation from the standard $\Lambda$CDM model leading to a higher value for $H_{0}$. Although the model considered in the previous section also deviates from the standard $\Lambda$CDM model, but according to the learned constraints, it is not clear if the $H_{0}$ tension can or not be solved. We will come to this in the next section. It is worth mentioning that according to these results, future observations with more lenses-sources systems and covering new redshift ranges will most likely not change the estimations of $\Omega^{(0)}_{dm}$. Moreover,  we observe from Fig. \ref{fig:Fig2_1}  that the evolution of the deceleration parameter $q$ and the evolution of $\Omega_{m}$ will not be strongly affected by future SLTD data measurements. This point can only be validated in the future. 

To finish this section, let us stress again that, according to BML by which SLTD data has been generated, the model considered can solve the $H_{0}$ tension and describes a quintessence dark energy dominated Universe where initially the dark energy is the cosmological constant.

\subsection{$f_{3}$CDM}

The third model we studied is $f_{3}$CDM which is given by Eqs.~(\ref{eq:f3}) and (\ref{eq:f31}). We similarly performed the generative process for BML using Eqs.~(\ref{TB1}) and (\ref{TB2}). Imposing the same flat priors as in the case of the $f_{2}$CDM model, we are able to learn the constraints on the model parameters. In particular, we have: 
\begin{itemize}
    \item When $z_{l}\in[0.1,1.2]$ and $z_{s}\in[0.3,1.7]$ for $N _{lens}= 50$ the best fit values with their $1\sigma$ errors are $\Omega^{(0)}_{dm} = 0.267 \pm 0.022$, $H_{0} = 67.58 \pm 0.161$  km/s/Mpc, $p = 5.16 \pm 0.1$, respectively. 
    
    \item On the other hand, when $z_{l}\in[0.1,1.2]$ and $z_{s}\in[0.3,1.7]$ with $N _{lens}= 100$ the best fit values and $1\sigma$ errors of the model parameters are found to be $\Omega^{(0)}_{dm} = 0.275 \pm 0.018$, $H_{0} = 73.44 \pm 0.1$  km/s/Mpc and $p = 5.15 \pm 0.1$.
    
    \item Moreover, for the other two cases, when $z_{l}\in[0.1,1.5]$ and $z_{s}\in[0.3,1.7]$, and $z_{l}\in[0.1,2.0]$ and $z_{s}\in[0.3,2.5]$ (in both cases $N _{lens}= 100$), we find that $\Omega^{(0)}_{dm} = 0.237 \pm 0.0167$, $H_{0} = 73.18 \pm 0.16$  km/s/Mpc and $p = 4.67 \pm 0.142$ $\Omega^{(0)}_{dm} = 0.259 \pm 0.01$, $H_{0} = 69.86 \pm 0.125$  km/s/Mpc and $p = 4.87 \pm 0.135$, respectively.
    
    \item Finally, for the case when $N _{lens}= 100$, we find $\Omega^{(0)}_{dm} = 0.256 \pm 0.006$, $H_{0} = 73.054 \pm 0.102$  km/s/Mpc and $p = 5.61 \pm 0.095$. In this case,  $z_{l}\in[0.1,2.4]$ and $z_{s}\in[0.3,2.5]$. 
\end{itemize}

\begin{table}[h!]
    \centering
    \begin{tabular}{ | c | c | c | c |  c |  c |  p{2cm} |}
    \hline
   
  $f_{3}$CDM: $f(T) = -\alpha T\left(1-e^{p\sqrt{T_{0}/T}}\right)$ & $N_{lens}$ & $\Omega^{(0)}_{dm}$ & $H_{0}$ & $p$  \\
      \hline
      
 when  $z_{l}\in[0.1,1.2]$ and $z_{s}\in[0.3,1.7]$ & $50$ & $0.267 \pm 0.022$ & $67.58 \pm 0.161$ & $5.16 \pm 0.1$\\
          \hline
          
when  $z_{l}\in[0.1,1.2]$ and $z_{s}\in[0.3,1.7]$  & $100$ & $ 0.275 \pm 0.018$ & $73.44 \pm 0.1$ & $5.15 \pm 0.1$ \\
         \hline

when $z_{l}\in[0.1,1.5]$ and $z_{s}\in[0.3,1.7]$ & $100$ & $0.237 \pm 0.0167$  & $73.18 \pm 0.16$ & $4.67 \pm 0.142$ \\
         \hline
         
 when $z_{l}\in[0.1,2.0]$ and $z_{s}\in[0.3,2.5]$ &  $100$  & $0.259 \pm 0.01$ & $69.86 \pm 0.125$ & $4.87 \pm 0.135$ \\  
 
     \hline 
     
 when $z_{l}\in[0.1,2.4]$ and $z_{s}\in[0.3,2.5]$ &  $100$  & $0.256 \pm 0.006$ & $73.054 \pm 0.102$ & $5.61 \pm 0.095$ \\  
 
     \hline      
    
    \end{tabular}
    \caption{Best fit values and $1\sigma$ errors estimated for $f_{3}$CDM given by Eqs.~(\ref{eq:f3}) and (\ref{eq:f31}), when the model based generation process is based on Eqs.~(\ref{TB1}) and (\ref{TB2}). The flat priors as $H_{0} \in [64,78]$, $\Omega^{(0)}_{dm} \in [0.2,0.4]$ and $p \in [-10,10]$ have been imposed, respectively. Note that $H_{0}$ is measured in the units of $ \mathrm{km} / \mathrm{s} / \mathrm{Mpc}$.}
    \label{tab:Table4}
\end{table}

\begin{figure}[h!]
\centering
\includegraphics[width=95 mm]{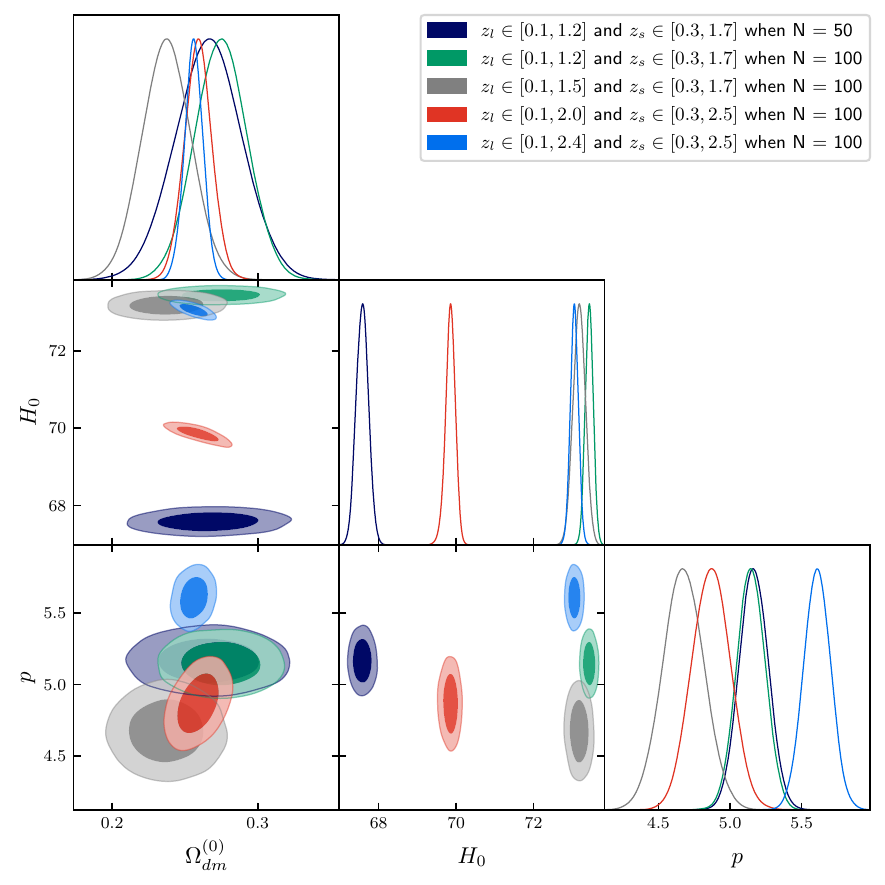}
\caption{The $1\sigma$ and $2\sigma$ confidence-level contour plots for the $f_{3}$CDM model, using the  of SLTD simulated data  obtained from the generative process based on Eqs.~(\ref{TB1}) and (\ref{TB2}). Each contour color stands for lenses and sources distributed over a specific redshift range with lens number $N_{\text{lens}}=50$ (navy contour) and $N_{\text{lens}}=100$ (green, gray, red, and blue contours), respectively. The flat priors as $H_{0} \in [64,78]$, $\Omega^{(0)}_{dm} \in [0.2,0.4]$ and $p \in [-10,10]$ have been imposed, respectively. Our initial belief used as an input is the $\Lambda$CDM model.}
\label{fig:Fig3_0}
\end{figure}

\begin{figure}[htb!]
 \centering
     \begin{subfigure}[b]{0.44\textwidth}
         \centering
         \includegraphics[width=\textwidth]{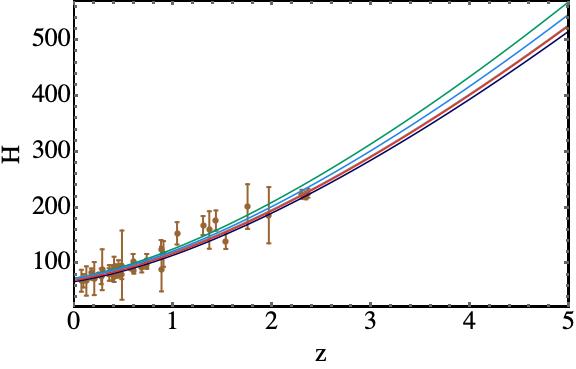}
         \caption{}
         \label{}
     \end{subfigure}
     \hskip -1exl
     \begin{subfigure}[b]{0.44\textwidth}
         \centering
         \includegraphics[width=\textwidth]{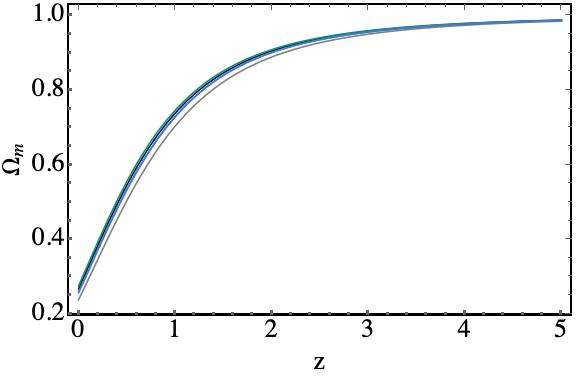}
         \caption{}
         \label{}
     \end{subfigure}
     \hskip -1ex
     \begin{subfigure}[b]{0.44\textwidth}
         \centering
         \includegraphics[width=\textwidth]{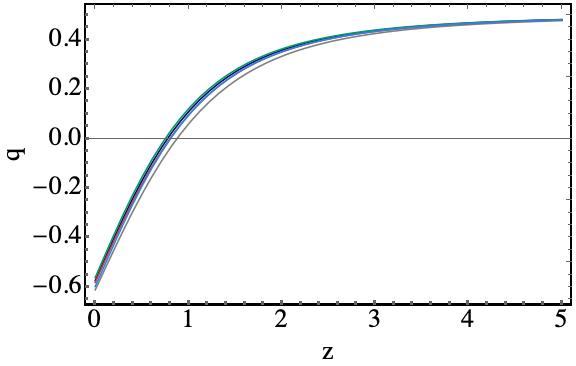}
         \caption{}
         \label{}
     \end{subfigure}
         \hskip -1ex
     \begin{subfigure}[b]{0.44\textwidth}
         \centering
         \includegraphics[width=\textwidth]{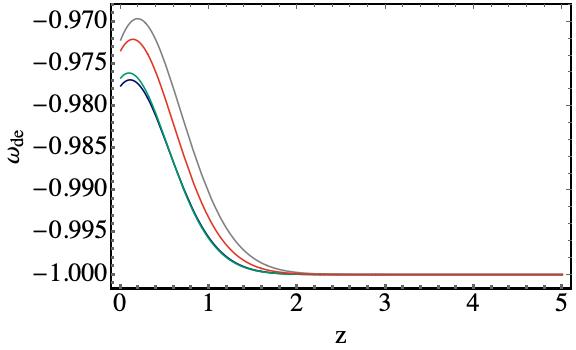}
         \caption{}
         \label{}
     \end{subfigure}
\caption{BML predictions for the redshift evolution of the Hubble parameter $H$, dark matter abundance $\Omega_{m}$, deceleration parameter $q$,  and equation of state parameter $\omega_{de} = \frac{p_{de}}{\rho_{de}}$, for the best fit values of the model parameters of $f_{3}$CDM presented in Table
\ref{tab:Table4}. Each  color line stands for lenses and sources distributed over a specific redshift range with lens number $N_{\text{lens}}=50$ (navy curve) and $N_{\text{lens}}=100$ (green, gray, red, and blue curves), respectively.}
 \label{fig:Fig3_1}
\end{figure}

The learned constraints on the model parameters have been summarized in Table \ref{tab:Table4} and the validation of the BML results for this model is presented in Fig. \ref{fig:Fig3_1}. We keep the same convention for the navy, green, grey, red, and blue curves as the legends in Fig. \ref{fig:Fig3_0} which shows the learned contour plots. We also note that the $b=1/p$ parameter having a larger value clearly indicates a deviation from the $\Lambda$CDM model.

From panel (a) of Fig. \ref{fig:Fig3_0} we note that the BML prediction for the redshift evolution of the Hubble function $H$ fits the observational data at low redshifts, though some tension can be observed at high redshifts. Moreover, panel (d) shows that the model behaves as the cosmological constant, $\omega_{de}=-1$, at $z \geq 2.0$ and the quintessence dark energy, $\omega_{de}>-1$, at $z<2.0$. In other words, the SLTD data-based learning shows that the recent Universe should contain quintessence dark energy, which started its evolution with a cosmological constant. In panel (c), a good phase transition between a decelerating and an accelerating phase can be noticed in the learned behaviour of the deceleration parameter $q$ for the five cases that we have considered in this paper. However, this model provided results different from the previous two models. In particular, the model can be used to solve the $H_{0}$ tension; however, one should note that the SLTD data is not able to give a final answer whether the model can solve the tension or not. In other words, the STLD data can strongly affect our understanding of how to solve the $H_{0}$ tension in $f(T)$ gravity. Moreover, it can strongly affect the constraints on $\Omega^{(0)}_{dm}$, indicating a tension there, too. This is another important consequence that BML allowed inferring from the study of this model.  

To end this section,we should mention that the different nature of BML as a tool combined with Eqs.~(\ref{eq:f3}) and (\ref{eq:f31}) allows us to develop a pipeline to study and constrain $f(T)$ gravity for cosmological purposes. It allows us to predict and learn how the $H_{0}$ tension can be solved and how the SLTD data can challenge it in $f(T)$ gravity. 

\section{Conclusions}\label{conclusions}

Using BML we have addressed the annoying $H_{0}$ tension. The real source of this issue is still unclear. A fair number of the attempts at solving the problem are based on the idea that the $H_{0}$ tension is not a mere statistical mismatch or artefact but that it is actually related to physical considerations. However, it must be mentioned that, despite very serious attempts to identify how this challenges our understanding of the Universe, there is still no reliable hint on the actual origin of the problem, and much work should be done yet. 

On the other hand, we may need to challenge the $\Lambda$CDM model to understand this issue. This can be done by challenging not only our understanding of dark energy but also the dark matter part. To this point, recently, it has been demonstrated using BML that there is a deviation from the cold dark matter paradigm on cosmological scales, which might efficiently solve the $H_{0}$ tension \cite{H06}. 

In the present paper, we have used BML to constrain $f(T)$ gravity-based cosmological models to see how the problem may be solved there. We have considered and learned the constraints on power-law, exponential, and square-root exponential $f(T)$ models using the SLTD as the main element for the generative process and the key ingredient of the Probabilistic ML approach. In this analysis, we did not rely on the lensing model itself, and what we needed is the redshifts of the lens and sources only. We would like to stress that very tight constraints on the parameters determining the $f(T)$ models have been obtained.

Moreover, our results contain a hint showing that more precise time delay measurements and the number of lensed systems could significantly affect the constraints on the model parameters. Taking into account the $H_{0}$ tension, we have validated the learned results with the available OHD and found a sign of tension that could exist between lensed GW+EM signals and OHD. Therefore, it is not excluded that utilizing both could lead to some misleading results in the model analysis. On the other hand, we have learned that exponential $f(T)$ in the light of SLTD data could solve the $H_{0}$ tension, while the power-law model slightly differs from the $\Lambda$CDM and definitely cannot solve the problem. 

The case of the power-law model is interesting for two reasons: first, we have learned that it could be very close to the $\Lambda$CDM model. Second, future SLTD data may indicate slight deviations from the $\Lambda$CDM model. In our opinion, this is another hint that in order to solve the $H_{0}$ tension, the $\Lambda$CDM model should be challenged. According to the learned best fit values of the parameters in the case of square-root exponential and exponential $f(T)$ models, we will have a quintessence dark energy dominated recent Universe, which for relatively high redshift evaluations contains a cosmological constant as dark energy.Although the alleviation of $H_0$ tension by late time modification is achieved through a phantom dark energy behavior\cite{Phantom} but recent studies suggest the tension  also can be resolved by reducing the ratio between effective Newton's constant $G_{eff}/G_{N}$ and Newton's constant to less than one\cite{Geff1, Geff2}. The exponential and squared exponential model in our analysis should therefore satisfy the condition $G_{eff}/G_n<1$ which leads to a faster $H_{0}$ expansion rate as we see in our analysis is the case. On the other hand, it has been shown that these models are statistically indistinguishable from $\Lambda$CDM model \cite{distin}, however,in our present work we found that exponential and squared exponential model deviate from $\Lambda$CDM model significantly as the value of their $p$ parameter ranging between 5 to 6. Moreover,the learned results indicate that an accelerated expanding phase transition will be observed naturally and smoothly in all three cases considered.

Finally, We would like to mention some interesting aspects that follow from our approach. Since the time delay distances can be measured from the lensed gravitational wave signals and their corresponding electromagnetic wave counterpart, our approach could be very useful during the source identification process. Indeed, we found a clear hint that we can have very strong constraints on lensed GW+EM systems and a reasonable combination of it with the simulations based on LSST, Einstein Telescope(ET), and the Dark Energy Survey(DES) can provide a powerful tool for the present cosmological analysis. A more detailed discussion of such possibilities will be the subject of a forthcoming paper. A final consideration is that the approach proposed in the present one can be easily extended to constrain the lensing systems.

\section*{Acknowledgements}
This work has been partially supported by MICINN (Spain), project PID2019-104397GB-I00, of the Spanish State Research
Agency program AEI/10.13039/501100011033, by the Catalan Government, AGAUR project 2017-SGR-247, and by the program Unidad de Excelencia Mar´ıa de Maeztu CEX2020-001058-M. This paper was supported by the Ministry of Education and Science of the Republic of Kazakhstan, grant AP08052034. MK has been supported by the Juan de la Cierva-incorporaci´on grant (IJC2020-042690-I).


\begin{thebibliography}{}
\bibitem{P1} 
H.~E.~S.~Velten, R.~F.~vom Marttens and W.~Zimdahl,
``Aspects of the cosmological \textquotedblleft{}coincidence problem\textquotedblright{},''
{\hypersetup{urlcolor=blue}\href{https://doi.org/110.1140/epjc/s10052-014-3160-4}{Phys. Rept. \textbf{513} (2012), 1-189}}, \href{https://arxiv.org/1410.2509}{arXiv:1410.2509 [astro-ph.CO]}.

\bibitem{P2} 
P.J. Steinhardt, \textit{in Critical Problems in Physics}, Edited by V. L. Fitch, D. R. Marlow, and M. A.E. Dementi (Princeton University Press, Princeton, 1997).




\bibitem{LCDM1} 
T.~Clifton, P.~G.~Ferreira, A.~Padilla and C.~Skordis,
``Modified Gravity and Cosmology,''{\hypersetup{urlcolor=blue}\href{https://doi.org/10.1016/j.physrep.2012.01.001}{Phys. Rept. \textbf{513} (2012), 1-189}}, \href{https://arxiv.org/1106.2476}{arXiv:1106.2476 [astro-ph.CO]}.




\bibitem{LCDM2}
L.~Baudis,
``Dark matter detection,''
{\hypersetup{urlcolor=blue}\href{https://doi.org/10.1088/0954-3899/43/4/044001}{J. Phys. G \textbf{43} (2016) no.4, 044001}}.

\bibitem{LCDM3}
G.~Bertone, D.~Hooper and J.~Silk,
``Particle dark matter: Evidence, candidates and constraints,''
{\hypersetup{urlcolor=blue}\href{https://doi.org/10.1016/j.physrep.2004.08.031}{Phys. Rept. \textbf{405} (2005), 279-390}}, \href{https://arxiv.org/hep-ph/0404175}{arXiv: 0404175 [hep-ph]}.

\bibitem{DE_start}
K.~Bamba, S.~Capozziello, S.~Nojiri and S.~D.~Odintsov,
``Dark energy cosmology: the equivalent description via different theoretical models and cosmography tests,''
{\hypersetup{urlcolor=blue}\href{https://doi.org/10.1007/s10509-012-1181-8}{Astrophys. Space Sci. \textbf{342} (2012), 155-228}}, \href{https://arxiv.org/1205.3421}{arXiv:1205.3421 [gr-qc]}.



\bibitem{DE_1}
C.~Li, X.~Ren, M.~Khurshudyan and Y.~F.~Cai,
``Implications of the possible 21-cm line excess at cosmic dawn on dynamics of interacting dark energy,''
{\hypersetup{urlcolor=blue}\href{https://doi.org/10.1016/j.physletb.2019.135141}{Phys. Lett. B \textbf{801} (2020), 135141}}, \href{https://arxiv.org/abs/1904.02458}{arXiv:1904.02458 [astro-ph.CO]}.


\bibitem{DE_2}
M.~Khurshudyan and R.~Myrzakulov,
``Phase space analysis of some interacting Chaplygin gas models,''
{\hypersetup{urlcolor=blue}\href{https://doi.org/10.1140/epjc/s10052-017-4634-y}{Eur. Phys. J. C \textbf{77} (2017) no.2, 65}}, \href{https://arxiv.org/abs/1509.02263}{arXiv:1509.02263 [gr-qc]}.

\bibitem{DE_3}
E.~Elizalde and M.~Khurshudyan,
``Cosmology with an interacting van der Waals fluid,''
{\hypersetup{urlcolor=blue}\href{https://doi.org/0.1142/S0218271818500372}{Int. J. Mod. Phys. D \textbf{27} (2017) no.04, 1850037}}, \href{https://arxiv.org/abs/1711.01143}{1711.01143 [gr-qc]}.



\bibitem{DE_4}
S.~Nojiri, S.~D.~Odintsov and T.~Paul,
``Barrow entropic dark energy: A member of generalized holographic dark energy family,''
{\hypersetup{urlcolor=blue}\href{https://doi.org/10.1016/j.physletb.2021.136844}{Phys. Lett. B \textbf{825} (2022), 136844}}, \href{https://arxiv.org/abs/2112.10159}{arXiv:2112.10159 [gr-qc]}.


\bibitem{DE_end}
S.~D.~Odintsov, D.~Saez-Chillon Gomez and G.~S.~Sharov,
``Testing the equation of state for viscous dark energy,''
{\hypersetup{urlcolor=blue}\href{https://doi.org/10.1103/PhysRevD.101.044010}{Phys. Rev. D \textbf{101} (2020) no.4, 044010}}, \href{https://arxiv.org/abs/2001.07945}{arXiv:2001.07945 [gr-qc]}.


\bibitem{Sol1} 
E.~J.~Copeland, M.~Sami and S.~Tsujikawa,
``Dynamics of dark energy,''
{\hypersetup{urlcolor=blue}\href{https://www.doi.org/10.1142/S021827180600942X
}{Int. J. Mod. Phys. D \textbf{15} (2006), 1753-1936}}, \href{https://arxiv.org/abs/hep-th/0603057 [hep-th]}{arXiv:0603057 [hep-th]}.

\bibitem{Sol2} 
Y.~F.~Cai, E.~N.~Saridakis, M.~R.~Setare and J.~Q.~Xia,
``Quintom Cosmology: Theoretical implications and observations,''
{\hypersetup{urlcolor=blue}\href{https://www.doi.org/10.1016/j.physrep.2010.04.001
}{Phys. Rept. \textbf{493} (2010), 1-60}}, \href{https://arxiv.org/abs/0909.2776}{arXiv:0909.2776 [hep-th]}.

\bibitem{F(R)1}
S.~Nojiri and S.~D.~Odintsov,
``Introduction to modified gravity and gravitational alternative for dark energy,''
{\hypersetup{urlcolor=blue}\href{https://www.doi.org/10.1142/S0219887807001928
}{eConf \textbf{C0602061} (2006), 06}}, \href{https://arxiv.org/abs/hep-th/0601213}{arXiv:0601213 [hep-th]}.



\bibitem{F(R)2}
W.~Hu and I.~Sawicki,
``Models of f(R) Cosmic Acceleration that Evade Solar-System Tests,''
{\hypersetup{urlcolor=blue}\href{https://www.doi.org/10.1103/PhysRevD.76.064004}{Phys. Rev. D \textbf{76} (2007), 064004}}, \href{https://arxiv.org/abs/0705.1158}{arXiv:0705.1158 [astro-ph]}



\bibitem{F(R)3}
S.~A.~Appleby and R.~A.~Battye,
``Do consistent $F(R)$ models mimic General Relativity plus $\Lambda$?,''
{\hypersetup{urlcolor=blue}\href{https://www.doi.org/10.1016/j.physletb.2007.08.037}{Phys. Lett. B \textbf{654} (2007), 7-12}}, \href{https://arxiv.org/abs/0705.3199}{arXiv:0705.3199 [astro-ph]}


\bibitem{F(R)4}
A.~A.~Starobinsky,
``Disappearing cosmological constant in f(R) gravity,''
{\hypersetup{urlcolor=blue}\href{https://www.doi.org/10.1134/S0021364007150027}{JETP Lett. \textbf{86} (2007), 157-163}}, \href{https://arxiv.org/abs/0706.2041}{arXiv:0706.2041 [astro-ph]}


\bibitem{F(R)5}
L.~Amendola, R.~Gannouji, D.~Polarski and S.~Tsujikawa,
``Conditions for the cosmological viability of f(R) dark energy models,''
{\hypersetup{urlcolor=blue}\href{https://www.doi.org/}{Phys. Rev. D \textbf{75} (2007), 083504}}, \href{https://arxiv.org/abs/gr-qc/0612180}{arXiv:0612180 [gr-qc]}



\bibitem{F(R)6}
G.~Cognola, E.~Elizalde, S.~Nojiri, S.~D.~Odintsov, L.~Sebastiani and S.~Zerbini,
`A Class of viable modified f(R) gravities describing inflation and the onset of accelerated expansion,''
{\hypersetup{urlcolor=blue}\href{https://doi.org/0.1103/PhysRevD.77.046009}{Phys. Rev. D \textbf{77} (2008), 046009}}, \href{https://arxiv.org/abs/0712.4017}{arXiv:0712.4017 [hep-th]}.



\bibitem{F(R)7}
T.~P.~Sotiriou and V.~Faraoni,
``f(R) Theories Of Gravity,''
{\hypersetup{urlcolor=blue}\href{https://www.doi.org/10.1103/RevModPhys.82.451
}{Rev. Mod. Phys. \textbf{82} (2010), 451-497}}, \href{https://arxiv.org/abs/0805.1726}{arXiv:0805.1726 [gr-qc]}.


\bibitem{F(R)8}
S.~Nojiri and S.~D.~Odintsov,
``Unified cosmic history in modified gravity: from F(R) theory to Lorentz non-invariant models,''
{\hypersetup{urlcolor=blue}\href{https://www.doi.org/10.1016/j.physrep.2011.04.001
}{Phys. Rept. \textbf{505} (2011), 59-144}}, \href{https://arxiv.org/abs/1011.0544}{arXiv:1011.0544 [gr-qc]}.



\bibitem{FT_3}
Y.~F.~Cai, S.~Capozziello, M.~De Laurentis and E.~N.~Saridakis,
``f(T) teleparallel gravity and cosmology,''
{\hypersetup{urlcolor=blue}\href{https://www.doi.org/10.1088/0034-4885/79/10/106901
}{Rept. Prog. Phys. \textbf{79} (2016) no.10, 106901}}, \href{https://arxiv.org/abs/1511.07586}{arXiv:1511.07586 [gr-qc]}.

\bibitem{AP1}
Y.~F.~Cai, S.~H.~Chen, J.~B.~Dent, S.~Dutta and E.~N.~Saridakis,
``Matter Bounce Cosmology with the f(T) Gravity,''
{\hypersetup{urlcolor=blue}\href{https://www.doi.org/10.1088/0264-9381/28/21/215011
}{Class. Quant. Grav. \textbf{28} (2011), 215011}}, \href{https://arxiv.org/abs/1104.4349}{arXiv:1104.4349 [astro-ph.CO]}.


\bibitem{AP2}
J.~B.~Dent, S.~Dutta and E.~N.~Saridakis,
``f(T) gravity mimicking dynamical dark energy. Background and perturbation analysis,''
{\hypersetup{urlcolor=blue}\href{https://www.doi.org/10.1088/1475-7516/2011/01/009
}{JCAP \textbf{01} (2011), 009}}, \href{https://arxiv.org/abs/1010.2215}{arXiv:1010.2215 [astro-ph.CO]}.

\bibitem{AP3}
R.~Myrzakulov,
``Accelerating universe from F(T) gravity,''
{\hypersetup{urlcolor=blue}\href{https://www.doi.org/10.1140/epjc/s10052-011-1752-9
}{Eur. Phys. J. C \textbf{71} (2011), 1752}}, \href{https://arxiv.org/abs/1006.1120}{arXiv:1006.1120 [gr-qc]}.

\bibitem{AP4}
R.~Zheng and Q.~G.~Huang,
``Growth factor in $f(T)$ gravity,''
{\hypersetup{urlcolor=blue}\href{https://www.doi.org/10.1088/1475-7516/2011/03/002
}{JCAP \textbf{03} (2011), 002}}, \href{https://arxiv.org/abs/1010.3512}{arXiv:arXiv:1010.3512 [gr-qc]}.

\bibitem{AP5}
K.~Bamba, G.~G.~L.~Nashed, W.~El Hanafy and S.~K.~Ibraheem,
``Bounce inflation in $f(T)$ Cosmology: A unified inflaton-quintessence field,''
{\hypersetup{urlcolor=blue}\href{https://www.doi.org/10.1103/PhysRevD.94.083513
}{Phys. Rev. D \textbf{94} (2016) no.8, 083513}}, \href{https://arxiv.org/abs/1604.07604}{arXiv:1604.07604 [gr-qc]}.

\bibitem{AP6}
K.~Bamba, S.~D.~Odintsov and E.~N.~Saridakis,
``Inflationary cosmology in unimodular $F(T)$ gravity,''
{\hypersetup{urlcolor=blue}\href{https://www.doi.org/10.1142/S0217732317501140
}{Mod. Phys. Lett. A \textbf{32} (2017) no.21, 1750114}}, \href{https://arxiv.org/abs/1605.02461}{arXiv:1605.02461 [gr-qc]}.

\bibitem{AP7}
S.~Carloni, F.~S.~N.~Lobo, G.~Otalora and E.~N.~Saridakis,
``Dynamical system analysis for a nonminimal torsion-matter coupled gravity,''
{\hypersetup{urlcolor=blue}\href{https://www.doi.org/10.1103/PhysRevD.93.024034
}{Phys. Rev. D \textbf{93} (2016), 024034}}, \href{https://arxiv.org/abs/1512.06996}{arXiv:1512.06996 [gr-qc]}.

\bibitem{AP8}
M.~Hohmann, L.~Jarv and U.~Ualikhanova,
``Dynamical systems approach and generic properties of $f(T)$ cosmology,''
{\hypersetup{urlcolor=blue}\href{https://www.doi.org/10.1103/PhysRevD.96.043508
}{Phys. Rev. D \textbf{96} (2017) no.4, 043508}}, \href{https://arxiv.org/abs/1706.02376}{arXiv:1706.02376 [gr-qc]}.

\bibitem{cons1}
R.~C.~Nunes, S.~Pan and E.~N.~Saridakis,
``New observational constraints on f(T) gravity from cosmic chronometers,''
{\hypersetup{urlcolor=blue}\href{https://www.doi.org/10.1088/1475-7516/2016/08/011
}{JCAP \textbf{08}, 011 (2016)}}, \href{https://arxiv.org/abs/1606.04359}{arXiv:1606.04359 [gr-qc]}.

\bibitem{cons2}
S.~Capozziello, G.~Lambiase and E.~N.~Saridakis,
``Constraining f(T) teleparallel gravity by Big Bang Nucleosynthesis,''
{\hypersetup{urlcolor=blue}\href{https://www.doi.org/10.1140/epjc/s10052-017-5143-8
}{Eur. Phys. J. C \textbf{77}, no.9, 576 (2017)}}, \href{https://arxiv.org/abs/1702.07952}{arXiv:1702.07952 [astro-ph.CO]}.

\bibitem{cons3}
R.~C.~Nunes,
``Structure formation in $f(T)$ gravity and a solution for $H_0$ tension,''
{\hypersetup{urlcolor=blue}\href{https://www.doi.org/10.1088/1475-7516/2018/05/052
}{JCAP \textbf{05}, 052 (2018)}}, \href{https://arxiv.org/abs/1802.02281}{arXiv:1802.02281 [gr-qc]}.


\bibitem{cons4}
M.~Benetti, S.~Capozziello and G.~Lambiase,
``Updating constraints on f(T) teleparallel cosmology and the consistency with Big Bang Nucleosynthesis,''
{\hypersetup{urlcolor=blue}\href{https://www.doi.org/10.1093/mnras/staa3368
}{Mon. Not. Roy. Astron. Soc. \textbf{500}, no.2, 1795-1805 (2020)}}, \href{https://arxiv.org/abs/2006.15335}{arXiv:2006.15335 [astro-ph.CO]}.

\bibitem{f1}
G.~R.~Bengochea and R.~Ferraro,
``Dark torsion as the cosmic speed-up,''
{\hypersetup{urlcolor=blue}\href{https://www.doi.org/10.1103/PhysRevD.79.124019
}{Phys. Rev. D \textbf{79}, 124019 (2009)}}, \href{https://arxiv.org/abs/0812.1205}{arXiv:0812.1205 [astro-ph]}.

\bibitem{f2}
K.~Bamba, C.~Q.~Geng, C.~C.~Lee and L.~W.~Luo,
``Equation of state for dark energy in $f(T)$ gravity,''
{\hypersetup{urlcolor=blue}\href{https://www.doi.org/10.1088/1475-7516/2011/01/021
}{JCAP \textbf{01}, 021 (2011)}}, \href{https://arxiv.org/abs/1011.0508}{arXiv:1011.0508 [astro-ph.CO]}.

\bibitem{f3}
X.~Ren, S.~F.~Yan, Y.~Zhao, Y.~F.~Cai and E.~N.~Saridakis,
``Gaussian processes and effective field theory of $f(T)$ gravity under the $H_0$ tension,''
\href{https://arxiv.org/abs/2203.01926}{arXiv:2203.01926 [astro-ph.CO]}.

\bibitem{f4}
X.~Ren, T.~H.~T.~Wong, Y.~F.~Cai and E.~N.~Saridakis,
``Data-driven Reconstruction of the Late-time Cosmic Acceleration with f(T) Gravity,''
{\hypersetup{urlcolor=blue}\href{https://www.doi.org/10.1016/j.dark.2021.100812
}{Phys. Dark Univ. \textbf{32} (2021), 100812}}, \href{https://arxiv.org/abs/2103.01260}{arXiv:2103.01260 [astro-ph.CO]}.


\bibitem{GP1}
Y.~F.~Cai, M.~Khurshudyan and E.~N.~Saridakis,
``Model-independent reconstruction of $f(T)$ gravity from Gaussian Processes,''
{\hypersetup{urlcolor=blue}\href{https://doi.org/10.3847/1538-4357/ab5a7f}{Astrophys. J. \textbf{888} (2020), 62}}, \href{https://arxiv.org/abs/1907.10813}{arXiv:1907.10813 [astro-ph.CO]}.

\bibitem{GP2}
E.~Elizalde and M.~Khurshudyan,
``Swampland criteria for a dark energy dominated universe ensuing from Gaussian processes and H(z) data analysis,''
{\hypersetup{urlcolor=blue}\href{https://doi.org/0.1103/PhysRevD.99.103533}{Phys. Rev. D \textbf{99} (2019) no.10, 103533}}, \href{https://arxiv.org/abs/1811.03861}{ArXiv:1811.03861 [astro-ph.CO]}.

\bibitem{GP3}
E.~\'O Colg\'ain and M.~M.~Sheikh-Jabbari,
``Elucidating cosmological model dependence with $H_0$,''
{\hypersetup{urlcolor=blue}\href{https://www.doi.org/10.1140/epjc/s10052-021-09708-2
}{Eur. Phys. J. C \textbf{81} (2021) no.10, 892}}, \href{https://arxiv.org/abs/2101.08565}{2101.08565 [astro-ph.CO]}.

\bibitem{P3}
E.~\'O.~Colg\'ain, M.~M.~Sheikh-Jabbari, R.~Solomon, G.~Bargiacchi, S.~Capozziello, M.~G.~Dainotti and D.~Stojkovic,
``Revealing Intrinsic Flat $\Lambda$CDM Biases with Standardizable Candles,''
\href{https://arxiv.org/abs2203.10558/}{arXiv:2203.10558 [astro-ph.CO]}.


\bibitem{H01}
N.~Aghanim \textit{et al.} [Planck],
``Planck 2018 results. VI. Cosmological parameters,''
{\hypersetup{urlcolor=blue}\href{https://www.doi.org/10.1051/0004-6361/201833910
}{Astron. Astrophys. \textbf{641} (2020), A6
[erratum: Astron. Astrophys. \textbf{652} (2021), C4]}}, \href{https://arxiv.org/abs/1807.06209}{arXiv:1807.06209 [astro-ph.CO]}.

\bibitem{H02}
A.~G.~Riess, S.~Casertano, W.~Yuan, J.~B.~Bowers, L.~Macri, J.~C.~Zinn and D.~Scolnic,
``Cosmic Distances Calibrated to 1\% Precision with Gaia EDR3 Parallaxes and Hubble Space Telescope Photometry of 75 Milky Way Cepheids Confirm Tension with $\Lambda$CDM,''
{\hypersetup{urlcolor=blue}\href{https://www.doi.org/10.3847/2041-8213/abdbaf
}{Astrophys. J. Lett. \textbf{908} (2021) no.1, L6}}, \href{https://arxiv.org/abs/2012.08534}{arXiv:2012.08534 [astro-ph.CO]}.

\bibitem{H03}
E.~Di Valentino, O.~Mena, S.~Pan, L.~Visinelli, W.~Yang, A.~Melchiorri, D.~F.~Mota, A.~G.~Riess and J.~Silk,
``In the realm of the Hubble tension\textemdash{}a review of solutions,''
{\hypersetup{urlcolor=blue}\href{https://www.doi.org/10.1088/1361-6382/ac086d
}{Class. Quant. Grav. \textbf{38} (2021) no.15, 153001}}, \href{https://arxiv.org/abs/2103.01183}{arXiv:2103.01183 [astro-ph.CO]}.

\bibitem{H04}
E.~Elizalde, M.~Khurshudyan, S.~D.~Odintsov and R.~Myrzakulov,
``Analysis of the $H_0$ tension problem in the Universe with viscous dark fluid,''
{\hypersetup{urlcolor=blue}\href{https://www.doi.org/10.1103/PhysRevD.102.123501
}{Phys. Rev. D \textbf{102} (2020) no.12, 123501}}, \href{https://arxiv.org/abs/2006.01879}{arXiv:2006.01879 [gr-qc]}.

\bibitem{H05}
E.~Elizalde and M.~Khurshudyan,
``Constraints on Cosmic Opacity from Bayesian Machine Learning: The hidden side of the $H_{0}$ tension problem,'' \href{https://arxiv.org/abs/2006.12913}{arXiv:2006.12913 [astro-ph.CO]}.


\bibitem{H06}
E.~Elizalde, J.~Gluza, M.~Khurshudyan,
``An approach to cold dark matter deviation and the $H_{0}$ tension problem by using machine learning,''
 \href{https://arxiv.org/abs/2104.01077}{arXiv:2104.01077 [astro-ph. C.O.]}.
 
 
\bibitem{H07}
D.~Wang and D.~Mota,
``Can $f(T)$ gravity resolve the $H_0$ tension?,''
{\hypersetup{urlcolor=blue}\href{https://www.doi.org/10.1103/PhysRevD.102.063530
}{Phys. Rev. D \textbf{102} (2020) no.6, 063530}}, \href{https://arxiv.org/abs/2003.10095}{2003.10095 [astro-ph.CO]}.


\bibitem{H08}
S.~D.~Odintsov and V.~K.~Oikonomou,
``Did the Universe experience a pressure non-crushing type cosmological singularity in the recent past?,''
{\hypersetup{urlcolor=blue}\href{https://www.doi.org/10.1209/0295-5075/ac52dc
}{EPL \textbf{137} (2022) no.3, 39001}}, \href{https://arxiv.org/abs/2201.07647}{arXiv:2201.07647 [gr-qc]}.

\bibitem{TD1}
T.~Treu,
``Strong Lensing by Galaxies,''
{\hypersetup{urlcolor=blue}\href{https://www.doi.org/10.1146/annurev-astro-081309-130924
}{Ann. Rev. Astron. Astrophys. \textbf{48}, 87-125 (2010)}}, \href{https://arxiv.org/abs/1003.5567}{arXiv:1003.5567 [astro-ph.CO]}.

\bibitem{HOLICOW}
S.~H.~Suyu, V.~Bonvin, F.~Courbin, C.~D.~Fassnacht, C.~E.~Rusu, D.~Sluse, T.~Treu, K.~C.~Wong, M.~W.~Auger and X.~Ding, \textit{et al.}
`H0LiCOW \textendash{} I. H0 Lenses in COSMOGRAIL's Wellspring: program overview,''
{\hypersetup{urlcolor=blue}\href{https://www.doi.org/10.1093/mnras/stx483
}{Mon. Not. Roy. Astron. Soc. \textbf{468} (2017) no.3, 2590-2604}}, \href{https://arxiv.org/abs/1607.00017}{arXiv:1607.00017 [astro-ph.CO]}

\bibitem{HOLICOW1}
K.~C.~Wong, S.~H.~Suyu, G.~C.~F.~Chen, C.~E.~Rusu, M.~Millon, D.~Sluse, V.~Bonvin, C.~D.~Fassnacht, S.~Taubenberger and M.~W.~Auger, \textit{et al.}
``H0LiCOW \textendash{} XIII. A 2.4 per cent measurement of H0 from lensed quasars: 5.3\ensuremath{\sigma} tension between early- and late-Universe probes,''
{\hypersetup{urlcolor=blue}\href{https://www.doi.org/10.1093/mnras/stz3094
}{Mon. Not. Roy. Astron. Soc. \textbf{498} (2020) no.1, 1420-1439}}, \href{https://arxiv.org/abs/1907.04869 }{arXiv:1907.04869 [astro-ph.CO]}.



\bibitem{GW1}
B.~P.~Abbott \textit{et al.} [LIGO Scientific, Virgo, 1M2H, Dark Energy Camera GW-E, DES, DLT40, Las Cumbres Observatory, VINROUGE and MASTER],
``A gravitational-wave standard siren measurement of the Hubble constant,''
{\hypersetup{urlcolor=blue}\href{https://www.doi.org/10.1038/nature24471
}{Nature \textbf{551} (2017) no.7678, 85-88}}, \href{https://arxiv.org/abs/1710.05835}{arXiv:1710.05835 [astro-ph.CO]}.


\bibitem{GW2}
B.~P.~Abbott \textit{et al.} [LIGO Scientific and Virgo],
``GW170817: Observation of Gravitational Waves from a Binary Neutron Star Inspiral,''
{\hypersetup{urlcolor=blue}\href{https://www.doi.org/10.1103/PhysRevLett.119.161101
}{Phys. Rev. Lett. \textbf{119} (2017) no.16, 161101}}, \href{https://arxiv.org/abs/1710.05832}{arXiv:1710.05832 [gr-qc]}.

\bibitem{GW3}
B.~P.~Abbott \textit{et al.} [LIGO Scientific, Virgo, Fermi-GBM and INTEGRAL],
``Gravitational Waves and Gamma-rays from a Binary Neutron Star Merger: GW170817 and GRB 170817A,''
{\hypersetup{urlcolor=blue}\href{https://www.doi.org/10.3847/2041-8213/aa920c
}{Astrophys. J. Lett. \textbf{848} (2017) no.2, L13}}, \href{https://arxiv.org/abs/1710.05834}{arXiv:1710.05834 [astro-ph.HE]}.

\bibitem{GW4}
D.~A.~Coulter, R.~J.~Foley, C.~D.~Kilpatrick, M.~R.~Drout, A.~L.~Piro, B.~J.~Shappee, M.~R.~Siebert, J.~D.~Simon, N.~Ulloa and D.~Kasen, \textit{et al.}
``Swope Supernova Survey 2017a (SSS17a), the Optical Counterpart to a Gravitational Wave Source,''
{\hypersetup{urlcolor=blue}\href{https://www.doi.org/10.1126/science.aap9811
}{Science \textbf{358} (2017), 1556}}, \href{https://arxiv.org/abs/1710.05452}{arXiv:1710.05452 [astro-ph.HE]}.

\bibitem{GW5}
K.~Hotokezaka, E.~Nakar, O.~Gottlieb, S.~Nissanke, K.~Masuda, G.~Hallinan, K.~P.~Mooley and A.~T.~Deller,
``A Hubble constant measurement from superluminal motion of the jet in GW170817,''
{\hypersetup{urlcolor=blue}\href{https://www.doi.org/10.1038/s41550-019-0820-1
}{Nature Astron. \textbf{3} (2019) no.10, 940-944}}, \href{https://arxiv.org/abs/1806.10596}{arXiv:1806.10596 [astro-ph.CO]}.

\bibitem{GW6}
H.~Y.~Chen, M.~Fishbach and D.~E.~Holz,
``A two per cent Hubble constant measurement from standard sirens within five years,''
{\hypersetup{urlcolor=blue}\href{https://www.doi.org/10.1038/s41586-018-0606-0
}{Nature \textbf{562} (2018) no.7728, 545-547}}, \href{https://arxiv.org/abs/1712.06531}{arXiv:1712.06531 [astro-ph.CO]}.


\bibitem{LSST1}
\v{Z}.~Ivezi\'c \textit{et al.} [LSST],
``LSST: from Science Drivers to Reference Design and Anticipated Data Products,''
{\hypersetup{urlcolor=blue}\href{https://www.doi.org/10.3847/1538-4357/ab042c
}{Astrophys. J. \textbf{873} (2019) no.2, 111}}, \href{https://arxiv.org/abs/0805.2366}{arXiv:0805.2366 [astro-ph]}.


\bibitem{LSST2}
M.~Oguri and P.~J.~Marshall,
``Gravitationally lensed quasars and supernovae in future wide-field optical imaging surveys,''
{\hypersetup{urlcolor=blue}\href{https://www.doi.org/10.1111/j.1365-2966.2010.16639.x
}{Mon. Not. Roy. Astron. Soc. \textbf{405} (2010), 2579-2593}}, \href{https://arxiv.org/abs/1001.2037}{arXiv:1001.2037 [astro-ph.CO]}

\bibitem{lens1}
P. Schneider, J. Ehlers, and E.E. Falco. \textit{Gravitational Lenses}. Springer, 1992.

\bibitem{lens2}
P. Schneider, C. S. Kochanek, and J. Wambsganss, \textit{Gravitational Lensing: Strong, Weak and Micro}. Springer, 2006.



\bibitem{K.L.}
S. Kullback and R.A. Leibler, ``On information and sufficiency,''
{\hypersetup{urlcolor=blue}\href{https://projecteuclid.org/journals/annals-of-mathematical-statistics/volume-22/issue-1/On-Information-and-Sufficiency/10.1214/aoms/1177729694.full}{Ann. Math. Stat., 22
(1951), 79-86}}.

\bibitem{ELBO}
D. M. Blei, A. Kucukelbir, J. D. McAuliffe,``Variational inference: A review for statisticians."
{\hypersetup{urlcolor=blue}\href{https://doi.org/10.48550/arXiv.1601.00670}{Journal of the American statistical Association 112, no. 518 (2017): 859-877.}}

\bibitem{51}
A. Graves, ``Practical variational inference for neural networks", \textit{In Advances in neural information processing systems}, pages 2348–2356, 2011.

\bibitem{52}
N. Metropolis, et al., ``Equation of state calculations by fast computing machines," 
{\hypersetup{urlcolor=blue}\href{https://doi.org/10.1063/1.1699114}{Journal of Chemical Physics 21 (1953): 1087-1092.}}


\bibitem{53}
J. Regier, A. C. Miller, D. Schlegel, R. P. Adams,
J. D. McAuliffe, and Prabhat, ``Approximate inference for
constructing astronomical catalogs from images",
\href{https://arxiv.org/abs/1803.00113}{arXiv:1803.00113 [stat. A.P.]}.
.

\bibitem{54}
G.~Gunapati, A.~Jain, P.~K.~Srijith and S.~Desai,
``Variational inference as an alternative to MCMC for parameter estimation and model selection,''
{\hypersetup{urlcolor=blue}\href{https://doi.org/10.1017/pasa.2021.64}{Publ. Astron. Soc. Austral. \textbf{39} (2022), e001}}, \href{https://arxiv.org/abs/1803.06473}{arXiv:1803.06473 [astro-ph.IM]}.




\bibitem{PyMC3}
J. Salvatier, T. Wiecki, C. Fonnesbeck, ``Probabilistic programming in Python using PyMC3"
{\hypersetup{urlcolor=blue}\href{
https://doi.org/10.48550/arXiv.1507.08050}{PeerJ Computer Science. 2016 Apr 6;2:e55}}, \href{https://arxiv.org/abs/1507.08050}{arXiv: 1507.08050 [stat.CO]}.
 




\bibitem{Phantom}
S.~Banerjee, M.~Petronikolou and E.~N.~Saridakis,
``Alleviating $H_0$ Tension with New Gravitational Scalar Tensor Theories,'' \href{https://arxiv.org/abs/2209.02426}{arXiv:2209.02426 [gr-qc]}.


\bibitem{Geff1}
L.~Kazantzidis and L.~Perivolaropoulos,
``\({\sigma }_{8}\) Tension. Is Gravity Getting Weaker at Low z? Observational Evidence and Theoretical Implications,''
{\hypersetup{urlcolor=blue}\href{https://www.doi.org/10.1007/978-3-030-83715-0\_33}{Modified Gravity and Cosmology. Springer, Cham.}}, \href{https://arxiv.org/abs/}{arXiv:1907.03176 [astro-ph.CO]}.


\bibitem{Geff2}
F.~K.~Anagnostopoulos, S.~Basilakos and E.~N.~Saridakis,
``Bayesian analysis of $f(T)$ gravity using $f\sigma_8$ data,''
{\hypersetup{urlcolor=blue}\href{https://www.doi.org/10.1103/PhysRevD.100.083517}{Phys. Rev. D \textbf{100} (2019) no.8, 083517}}, \href{https://arxiv.org/abs/1907.07533}{arXiv:1907.07533 [astro-ph.CO]}.


\bibitem{distin}
S.~Nesseris, S.~Basilakos, E.~N.~Saridakis and L.~Perivolaropoulos,
``Viable $f(T)$ models are practically indistinguishable from $\Lambda$CDM,''
{\hypersetup{urlcolor=blue}\href{https://www.doi.org/10.1103/PhysRevD.88.103010}{Phys. Rev. D \textbf{88} (2013), 103010}}, \href{https://arxiv.org/abs/1308.6142}{arXiv:1308.6142 [astro-ph.CO]}.

\end{thebibliography}
\end{document}